\documentclass[sigconf]{acmart}
\AtBeginDocument{%
  }
\usepackage{lipsum}  

\copyrightyear{2024}
\acmYear{2024}
\setcopyright{acmlicensed}
\acmConference[KDD '24] {Proceedings of the 30th ACM SIGKDD Conference on Knowledge Discovery and Data Mining }{August 25--29, 2024}{Barcelona, Spain.}
\acmBooktitle{Proceedings of the 30th ACM SIGKDD Conference on Knowledge Discovery and Data Mining (KDD '24), August 25--29, 2024, Barcelona, Spain}
\acmISBN{979-8-4007-0490-1/24/08}
\acmDOI{10.1145/XXXXXX.XXXXXX}

\usepackage{enumitem}
\usepackage{multirow}
\usepackage{graphicx}
\usepackage{stfloats}
\newcommand{\proposed}{\textsf{A-LLMRec}}
\begin{document}

% \title{Incorporating LLMs for the Utilization of Collaborative Knowledge in Sequential Recommendation}

\title{Large Language Models meet Collaborative Filtering: An Efficient All-round LLM-based Recommender System}

\author{Sein Kim}
\authornote{Both authors contributed equally to this research.}
\email{rlatpdlsgns@kaist.ac.kr}
\affiliation{
\institution{KAIST}
\country{Republic of Korea}
}

\author{Hongseok Kang}
\email{ghdtjr0311@kaist.ac.kr}
\authornotemark[1]
\affiliation{
\institution{KAIST}
\country{Republic of Korea}
}

\author{Seungyoon Choi}
\email{csyoon08@kaist.ac.kr}
\affiliation{
\institution{KAIST}
\country{Republic of Korea}
}

\author{Donghyun Kim}
\email{amandus.kim@navercorp.com}
\affiliation{
\institution{NAVER Corporation}
\country{Republic of Korea}
}

\author{Minchul Yang}
\email{minchul.yang@navercorp.com}
\affiliation{
\institution{NAVER Corporation}
\country{Republic of Korea}
}

\author{Chanyoung Park}
\authornote{Corresponding author.}
\email{cy.park@kaist.ac.kr}
\affiliation{
\institution{KAIST}
\country{Republic of Korea}
}

\begin{abstract}
% Collaborative filtering recommender systems (RecSys) play a crucial role in enhancing the user experience on social media and e-commerce platforms. 
Collaborative filtering recommender systems (CF-RecSys) have shown successive results in enhancing the user experience on social media and e-commerce platforms. However, as CF-RecSys struggles under cold scenarios with sparse user-item interactions, recent strategies have focused on leveraging modality information of user/items (e.g., text or images) based on pre-trained modality encoders and Large Language Models (LLMs).
% mitigate this by incorporating modality information and utilizing Large Language Models (LLMs).
% Despite their effectiveness in cold scenarios, they suffer from a lack of collaborative knowledge and underperform in warm scenarios with abundant interactions. 
Despite their effectiveness under cold scenarios, we observe that they underperform simple traditional collaborative filtering models under warm scenarios due to the lack of collaborative knowledge.
In this work, we propose an efficient \textbf{A}ll-round \textbf{LLM}-based \textbf{Rec}ommender system, called~\proposed, that excels not only in the cold scenario but also in the warm scenario.
Our main idea is to enable an LLM to directly leverage the collaborative knowledge contained in a pre-trained state-of-the-art CF-RecSys so that the emergent ability of the LLM as well as the high-quality user/item embeddings that are already trained by the state-of-the-art CF-RecSys can be jointly exploited.
This approach yields two advantages: (1) model-agnostic, allowing for integration with various existing CF-RecSys, and (2) efficiency, eliminating the extensive fine-tuning typically required for LLM-based recommenders. 
Our extensive experiments on various real-world datasets demonstrate the superiority of~\proposed~in various scenarios, including cold/warm, few-shot, cold user, and cross-domain scenarios. 
% \textcolor{blue}{~\proposed~ also propose the possibility of capability extension of CF-RecSys in generation tasks beyond item recommendation.}
Beyond the recommendation task, we also show the potential of~\proposed~in generating natural language outputs based on the understanding of the collaborative knowledge by performing a favorite genre prediction task.
% \proposed~also extends the capabilities of CF-RecSys beyond item recommendation, \textcolor{blue}{facilitating tasks such as recommendation explanation and item review generation}
Our code is available at \url{https://github.com/ghdtjr/A-LLMRec}.

% Addressing this challenge, we introduce LLM-based recommender ~\proposed~(Incorporating LLMs for the Utilization of Collaborative Knowledge in Sequential RecommendatION), designed to operate effectively across both cold and warm scenarios. ~\proposed~ stands out by harmoniously blending the collaborative knowledge inherent in RecSys with the advanced capabilities of LLMs via an innovative alignment network. This network aligns the item embeddings of RecSys with the token space of LLMs, enabling the transfer of collaborative knowledge without the need for fine-tuning either RecSys or LLMs. This approach yields two advantages: model-agnostic, allowing for integration with various existing RecSys, and cost-efficient, eliminating the extensive fine-tuning typically required for LLM-based recommenders. Our extensive experiments demonstrate ~\proposed's superior performance RecSys in various scenarios, including cold and warm, few-shot, and cold user scenarios. ~\proposed~ also extends the capabilities of RecSys beyond item recommendation, facilitating tasks such as recommendation explanation and item review generation.
\end{abstract}





\maketitle

\section{Introduction}
\looseness=-1
With the recent exponential growth in the number of users and items, collaborative filtering models \cite{he2017neural, sun2019bert4rec, kang2018self, he2020lightgcn} encounter the long-standing cold-start problem \cite{10.1145/3109859.3109912, NIPS2017_dbd22ba3}, stemming from the inherent sparsity of user-item interaction data. In other words, for users/items with few interactions, it becomes challenging to construct collaborative knowledge with other similar users/items, leading to suboptimal recommendation performance, especially in the cold-start scenarios. 
To overcome this issue, recent studies have focused on leveraging modality information of users/items (e.g., user demographics, item titles, descriptions, or images) to enhance recommendation performance under cold-start scenarios. Specifically, MoRec \cite{10.1145/3539618.3591932} utilizes pre-trained modality encoders (e.g., BERT~\cite{devlin-etal-2019-bert} or Vision-Transformer~\cite{dosovitskiy2021an}) to project raw modality features of items (e.g., item texts or images), thereby replacing the item embeddings typically used in collaborative filtering recommendation models. Similarly, CTRL \cite{li2023ctrl} considers tabular data and its textual representation as two different modalities and uses them to pre-train collaborative filtering recommendation models through a contrastive learning objective, which is then fine-tuned for specific recommendation tasks.

\begin{figure}[t]
    \centering
    \includegraphics[width=0.87\linewidth]{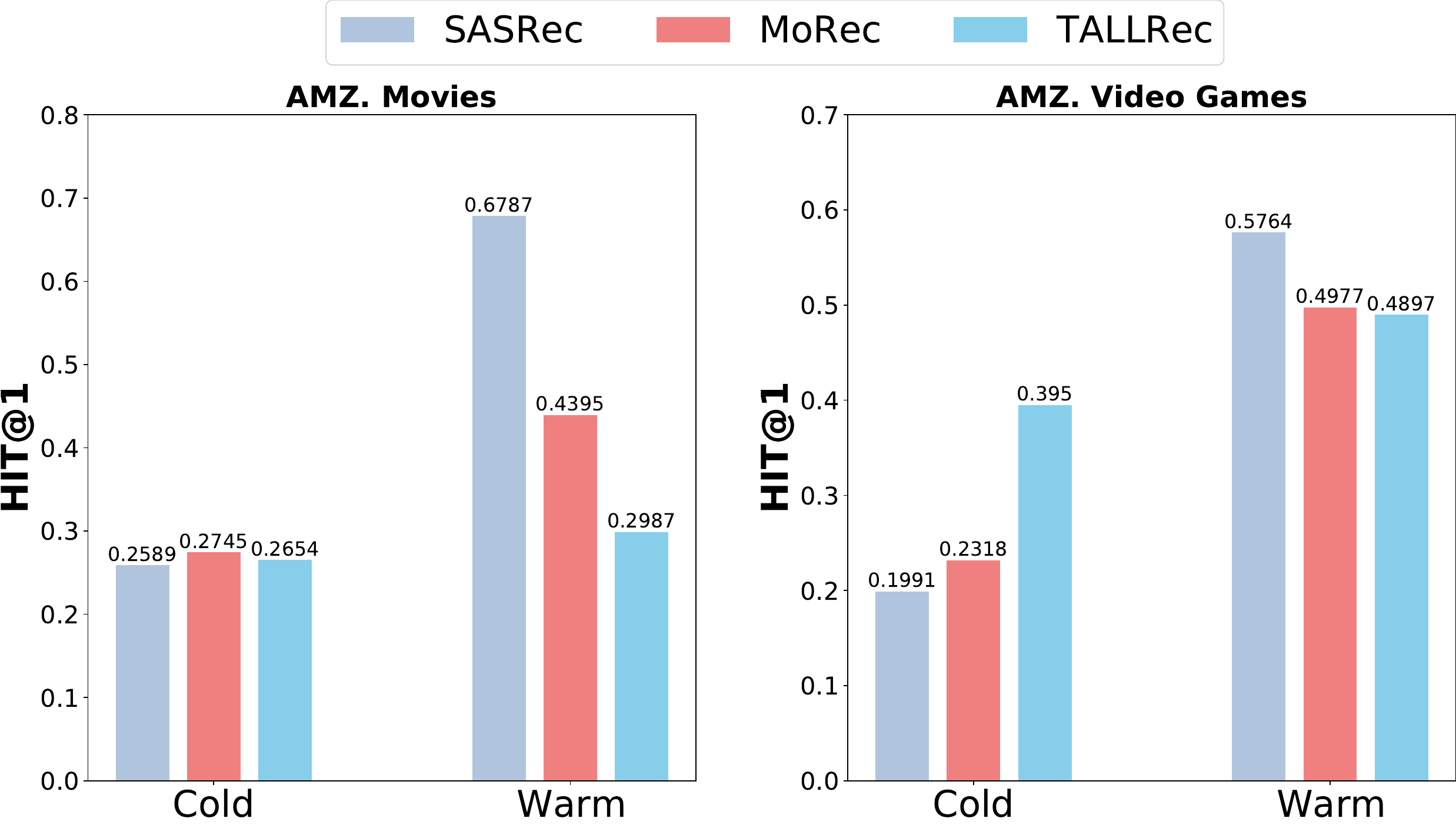}
    \vspace{-2ex}
    \caption[Caption for LOF]
    {Comparisons between collaborative filtering model (SASRec), modality-aware model (i.e., MoRec), and LLM-based model (i.e., TALLRec) under the cold/warm\footnotemark[1] scenarios on Amazon Movies/Video Games dataset (Hit@1)\footnotemark[2].}
    \label{fig: warm cold}
    \vspace{-4ex}
\end{figure}

\addtocounter{footnote}{1}
\footnotetext{An item is categorized as `warm' if it falls within the top 35\% of interactions, and if it falls within the bottom 35\%, it is classified as a `cold' item.}
\addtocounter{footnote}{1}
\footnotetext{After training each model using all the available data in the training set, we separately evaluate on cold and warm items in the test set.}

% LLM의 등장 + 관심 -> modality information을 capture 하는 방법론에 대한 제시
% Another strategy to tackle the cold start problem in recommendation systems is leveraging the Large Language Models (LLMs). 
Despite the effectiveness of modality-aware recommender systems in cold scenarios, the recent emergence of Large Language Models (LLMs), known for their rich pre-trained knowledge and advanced language understanding capabilities, has attracted significant interest in the recommendation domain to effectively extract and integrate modality information \cite{wu2023survey, sanner2023large}. 
% Specifically, based on pre-trained knowledge and advanced language comprehension, approaches have been proposed leveraging LLMs to capture the modality information. 
% More precisely, leveraging pre-trained knowledge and their advanced language understanding capabilities, recent approaches utilize LLMs to effectively extract and integrate modality information.
Early studies on LLM-based recommendation \cite{gao2023chat,wang2023zero, he2023large} have employed OpenAI-GPT with \textit{In-context Learning} \cite{brown2020language}. This approach adapts to new tasks or information based on the context provided within the input prompt and demonstrates the potential of LLMs as a recommender system. 
Moreover, to bridge the gap between the training tasks of LLMs and recommendation tasks, TALLRec \cite{bao2023tallrec} fine-tunes LLMs with recommendation data using LoRA \cite{hu2022lora}. 
% utilizes the \textit{Parameter Efficient Fine-Tuning} (PEFT) method, commonly referred to as LoRA \cite{hu2022lora}, to fine-tune the LLMs.
This approach has empirically demonstrated that, in cold scenarios and cross-domain scenarios, fine-tuned LLMs outperform traditional collaborative filtering models.
% So far various methods utilizing LLM have been proposed to resolve cold start and cross-domain challenges in recommendation. The pre-trained knowledge and reasoning power of LLM based on the advanced comprehension of language is well-suited for tackling cold start and cross-domain issues. [['Zero-shot conversational recommender Reference]] empirically demonstrated that LLMs, with carefully crafted prompts, outperformed other conversational recommendation models in a zero-shot manner. Additionally, [[ TALLRec Reference]] involved fine-tuning LLMs using the PEFT (Parameter Efficient Fine-Tuning) method, commonly known as LoRA. In a few-shot user setting, [[ TALLRec ]] manifested superior performance compared with the other conventional recommendation models.

% LLM4Rec의 한계점 (warm setting에서 기존 모델에서 지는 상황)
% CY: TaLLRec은 왜 finetune을 하는데도 collaborative정보다 안담기는지? rec task와 align을 하는데도? 만약 적은 개수만 사용해서 fine-tune해서 그런거면, 많은 데이터로 fine-tune하면 더 향상의 여지가 있나? -> need discussion

Although modality-aware and LLM-based recommender systems have proven effective in cold scenarios with limited user-item interactions, we argue that these methods suffer from the lack of collaborative knowledge due to their heavy reliance on textual information~\cite{10.1145/3539618.3591932}.
% [[Zero Shot Conversational Recommender]] demonstrated that LLMs conduct recommendations by utilizing item information based on text, rather than leveraging collaborative filtering knowledge.
% Specifically, existing LLM-based recommenders \cite{bao2023tallrec, li2023ctrl,wang2023zero} often overlook the importance of preserving and integrating the collaborative knowledge, resulting in models that fail to exploit collaborative filtering capabilities fully \cite{10.1145/3543507.3583434, 10.1145/3539618.3591932}. 
Consequently, \textit{when abundant user-item interactions are available (i.e., warm scenario), modality-aware and LLM-based recommenders are rather inferior to simple traditional collaborative filtering models}. 
% Figure \ref{fig: warm cold} shows the performance comparisons between a traditional collaborative filtering model (i.e., SASRec), a modality-aware recommendation model (i.e., MoRec), and an LLM-based recommendation model (i.e., TALLRec) under the cold and warm scenarios. 
% \textcolor{blue}{We trained each model using all available data and then conducted inference separately on both cold and warm items.}
As shown in Figure \ref{fig: warm cold}, while the modality-aware recommender (i.e., MoRec) and the LLM-based recommender (i.e., TALLRec) significantly outperform the traditional collaborative filtering model (i.e., SASRec~\cite{kang2018self}) in the cold scenario, they are outperformed by the traditional collaborative filtering model in the warm scenario. This is mainly because the textual information becomes less important in the warm scenario, where ID-based collaborative filtering models excel at modeling popular items~\cite{10.1145/3539618.3591932, 10.1145/3404835.3462919}. 
% It is important to note that since TaLLRec simply converts the conventional recommendation task into an instruction text and use it for fine-tuning, it still fails to explicitly capture the collaborative knowledge that is crucial in warm scenarios.
However, while excelling in the cold scenario is crucial, the majority of user interactions and the revenue are predominantly generated from already existing and active items (i.e., warm items) in real-world application of recommendation systems, which contribute up to 90\% of interactions in offline-industrial data \cite{10.1145/3539618.3591856, 09bbc5fe-8976-3a1a-8169-b0d39f201d67}. Furthermore, as demonstrated by DCBT \cite{10.1145/3539618.3591856}, modeling both warm and cold items is essential for improving overall user engagement, which is evidenced by A/B testing with real-world industrial data. This implies that the warm scenario should not be overlooked.

\looseness=-1
In this paper, we propose an efficient all-round LLM-based recommender system, called~\proposed~(\textbf{A}ll-round \textbf{LLM}-based \textbf{Rec}-ommender system), that excels not only in the cold scenario but also in the warm scenario (hence, all-round recommender system).
Our main idea is to enable an LLM to directly leverage the collaborative knowledge contained in a pre-trained state-of-the-art collaborative filtering recommender system (CF-RecSys) so that the emergent ability~\cite{wei2022emergent} of the LLM, as well as the high-quality user/item embeddings that are already trained by the state-of-the-art CF-RecSys, can be jointly exploited.
% whose goal is to exploit the emergent ability~\cite{wei2022emergent} of the LLM as well as the high-quality user/item embeddings that are already trained by the state-of-the-art RecSys.
More precisely, we devise an alignment network that aligns the item embeddings of the CF-RecSys with the token space of the LLM, aiming at transferring the collaborative knowledge learned from a pre-trained CF-RecSys to the LLM enabling it to understand and utilize the collaborative knowledge for the downstream
recommendation task. 

The key innovation of~\proposed~is that it \textit{requires the fine-tuning of neither the CF-RecSys nor the LLM}, and that the alignment network is the only neural network that is trained in~\proposed, which comes with the following two crucial advantages:
% \vspace{-2ex}
\begin{enumerate}[leftmargin=0.5cm]
    \item \textbf{(Model-agnostic)} \proposed~allows any existing CF-RecSys to be integrated, which implies that services using their own recommender models can readily utilize the power of the LLM. Besides, any updates of the recommender models can be easily reflected by simply replacing the old models, which makes the model practical in reality.
    \item \textbf{(Efficiency)} ~\proposed~is efficient in that the alignment network is the only trainable neural network, while TALLRec \cite{bao2023tallrec} requires the fine-tuning of the LLM with LoRA \cite{hu2022lora}. As a result, \proposed~trains approximately \textbf{2.53} times and inferences \textbf{1.71} times faster than TALLRec, while also outperforming both TALLRec and CF-RecSys in both cold and warm scenarios.
\end{enumerate}
Our extensive experiments on various real-world datasets demonstrate the superiority of~\proposed, revealing that aligning high-quality user/item embeddings with the token space of the LLM is the key for solving not only cold/warm scenarios but also few-shot, cold user, and cross-domain scenarios. 
Lastly, beyond the recommendation task, we perform a language generation task, i.e., favorite genre prediction, to demonstrate that~\proposed~can generate natural language outputs based on the understanding of users and items through the aligned collaborative knowledge from CF-RecSys.
% \textcolor{blue}{Furthermore, we propose the possibility that integrating collaborative knowledge into an LLM enables natural language tasks.}
% Furthermore, we show that integrating the collaborative knowledge into an LLM enables various natural language tasks including \textcolor{blue}{[Rewrite]recommendation explanation and item review generation. }
Our main contributions are summarized as follows:
\begin{itemize}[leftmargin=0.5cm]
\item We present an LLM-based recommender system, called~\proposed, that directly leverages the collaborative knowledge contained in a pre-trained state-of-the-art recommender system.
\item \proposed~requires the fine-tuning of neither the CF-RecSys nor the LLM, while only requiring an alignment network to be trained to bridge between them. 
\item Our extensive experiments demonstrate that \proposed~outperforms not only the conventional CF-RecSys in the warm scenario but also the LLMs in the cold scenario.
% Specifically, LLMs in~\proposed~can understand the modality information consisting collaborative and item descriptions. Consequently, we empirically demonstrate that ~\proposed outperforms conventional RecSys both in the warm and cold scenario, few-shot, and cold user scenario.
% \item We propose a time and cost-efficient method for constructing outperforming LLM-based recommendations without fine-tuning LLM or pre-trained RecSys.
% \item Thanks to ~\proposed 's training-free approach on RecSys and LLMs, we demonstrate the model-agnostic property of~\proposed, which enables application to any existing RecSys, facilitating the integration of LLM's power into the RecSys that companies already have.

% TODO: item, text align 어필, 큰 데이터셋에 적용
% \item In our methods, we verified the LLM-based recommendation in a huge dataset with a 290 thousands of users
% \item While the alignment of item space and language space, 
\end{itemize}

\section{Related Work}
\subsection{Collaborative Filtering}
Collaborative Filtering (CF) is the cornerstone of recommendation systems, fundamentally relying on leveraging users' historical preferences to inform future suggestions. The key idea is to rely on similar users/items for recommendations.
% It began with heuristic methods for computing similarities of items or users. 
The emergence of matrix factorization marked a significant advancement in CF, as evidenced by numerous studies \cite{sarwar2001item, koren2009matrix, hu2008collaborative}, demonstrating its superiority in capturing the latent factors underlying user preferences. This evolution continued with the introduction of Probabilistic Matrix Factorization (PMF) \cite{mnih2007probabilistic, chaney2015probabilistic} and Singular Value Decomposition (SVD) \cite{ma2008guide, zhou2015svd}, which integrate probabilistic and decomposition techniques to further refine the predictive capabilities of CF models. AutoRec \cite{sedhain2015autorec} and Neural Matrix Factorization (NMF) \cite{he2017neural} utilized deep learning to enhance CF by capturing complex user-item interaction patterns. Recently, \cite{10.1145/1772690.1772773, 10.5555/2540128.2540504, 10.1145/3580305.3599516, 10.1145/3583780.3614976} proposed modeling collaborative filtering based on sequential interaction history. Caser \cite{tang2018personalized} and NextItNet \cite{yuan2019simple} utilize Convolutional Neural Networks (CNNs) \cite{NIPS2012_c399862d} to capture the local sequence information, treating an item sequence as images. While these methods effectively capture user preferences using interaction history, including user and item IDs, they overlook the potential of the modality information of the user/item, which could enhance model performance and offer a deeper analysis of user behaviors.

\subsection{Modality-aware Recommender Systems}
% \textcolor{red}{(CY: Content here is almost exactly the same as what was written in the introduction (although I have rewritten the introduction). Related work section is meant to be a more detailed description of relevant literature, not just copying and pasting the introduction. You should mention about more studies of modality-aware RecSys beyond MoRec and CTRL. Same applies for 2.2 LLM-based RecSys. Let me know once you revised 2.1 and 2.2.)}
Modality-aware recommenders utilize modality information such as item titles, descriptions, or images to enhance the recommendation performance mainly under cold scenarios. Initially, CNNs were used to extract visual features, modeling human visual preferences based on Mahalanobis distance \cite{10.1145/2766462.2767755}. With advancements in pre-trained modality encoders like BERT \cite{devlin-etal-2019-bert, liu2021noninvasive, 10.1145/3539618.3591932, 10.1145/3343031.3351034, 10.1145/3512527.3531378} and ResNet/Vision-Transformer \cite{dosovitskiy2021an, 10.1145/3503161.3548405}, modality-aware recommender systems have accelerated research by utilizing modality knowledge on recommendation tasks. 
For example, NOVA \cite{liu2021noninvasive} and DMRL \cite{liu2022disentangled} proposed non-invasive fusion and disentangled fusion of modality, respectively, by carefully integrating pure item embeddings and text-integrated item embeddings using the attention mechanism. MoRec \cite{10.1145/3539618.3591932} leverages modality encoders to project raw modality features, thereby replacing item embeddings used in collaborative filtering models. As for the pre-training based models, \citet{10.1145/3512527.3531378} constructs user-user and item-item co-interaction graphs to extract collaborative knowledge, then integrates with user/item text information through attention mechanism in an auto-regressive manner, and CTRL \cite{li2023ctrl} pre-trains the collaborative filtering models using paired tabular data and textual data through a contrastive learning objective, subsequently fine-tuning them for recommendation tasks. 
Most recently, RECFORMER \cite{10.1145/3580305.3599519} proposed to model user preferences and item features as language representations based on the Transformer architecture by formulating the sequential recommendation task as the next item sentence prediction task, where the item key-value attributes are flattened into a sentence.
% However, this method is specifically tailored to the sequential recommendation task~\cite{hou2022towards,10.1145/3580305.3599519}, whereas~\proposed~can be applied to any type of tasks depending on the backbone CF-RecSys adopted.
% \textcolor{blue}{The RECFORMER \cite{10.1145/3580305.3599519} considers each item text sentence as a single item, converting user item sequence into sequence of sentences. Instead of using pre-trained modality encoders, RECFORMER trains Transformer \cite{vaswani2017attention} from scratch to understand and predict the next item sentence. This approach conflicts with leveraging pre-trained modality encoders to enhance recommendation performance, so we didn't use Recformer as modality-aware recommender in this study.}
% Recently, research on LLMs has gained prominence in the field of modality-aware recommendation systems, with LLM-based recommendations emerging as a significant area of focus.

\subsection{LLM-based Recommender Systems}
Recently, research on LLMs has gained prominence in the field of modality-aware recommendation systems, with LLM-based recommendations emerging as a significant area of focus.
% Recently, the emergence of LLM's remarkable generalization ability for unseen tasks and domains has attracted extensive attention. 
The pre-trained knowledge and the reasoning power of LLMs based on the advanced comprehension of language are shown to be effective for recommendation tasks, and many approaches have been proposed leveraging LLM as a recommender system. More precisely, \cite{gao2023chat,wang2023zero, he2023large} utilize LLMs with \textit{In-context Learning} \cite{brown2020language}, adapting to new tasks or information based on the context provided within the input prompt. 
For example, \citet{sanner2023large} employs {In-context Learning} for recommendation tasks, exploring various prompting styles such as completion, instructions, and few-shot prompts based on item texts and user descriptions. \citet{gao2023chat} assigns the role of a recommender expert to rank items that meet users' needs through prompting and conducts zero-shot recommendations. These studies empirically demonstrated the potential of LLMs using its rich item information and natural language understanding in the recommendation domain. However, these approaches often underperform traditional recommendation models \cite{sun2019bert4rec,kang2018self}, due to the gap between the natural language downstream tasks used for training LLMs and the recommendation task \cite{bao2023tallrec}. To bridge this gap, TALLRec \cite{bao2023tallrec} employs the \textit{Parameter Efficient Fine-Tuning} (PEFT) method, also known as LoRA \cite{hu2022lora}. This methodology enables TALLRec to demonstrate enhanced efficacy, surpassing traditional collaborative filtering recommendation models, particularly in mitigating the challenges posed by the cold start dilemma and in navigating the complexities of cross-domain recommendation scenarios.
However, it is important to note that since TALLRec simply converts the conventional recommendation task into an instruction text and uses it for fine-tuning, it still fails to explicitly capture the collaborative knowledge that is crucial in warm scenarios.
% Due to the powerful modality encoder, the models lose the collaborative knowledge which hinders the user-item interaction information and only concentrates on contextual information of items.

% \section{Preliminaries}
% In this section, we introduce a formal definition of the problem including the notations and the task description (Section \ref{Sec problem definition}) followed by the description of our prompt design (Section \ref{Sec Prompt Engineering}).

\section{Problem Formulation}\label{Sec problem definition}
In this section, we introduce a formal definition of the problem including the notations and the task description.

\smallskip
\noindent \textbf{Notations.} Let $\mathcal{D}$ denote the historical user-item interaction dataset $(\mathcal{U}, \mathcal{I}, \mathcal{T}, \mathcal{S}) \in \mathcal{D}$, where $\mathcal{U}, \mathcal{I}, \mathcal{T}, \text{and}~\mathcal{S}$ denote the set of users, items, item titles/descriptions, and item sequences, respectively. 
$\mathcal{S}^u = (i^u_1, i^u_2, \cdots, i^u_k, \cdots i^u_{|\mathcal{S}^u|}) \in \mathcal{S}$ is a sequence of item interactions of a user $u \in \mathcal{U}$, where $i^u_k$ denotes the $k$-th interaction of user $u$, and this corresponds to the index of the interacted item in the item set $\mathcal{I}$.
% $\mathcal{S}^u \in \mathcal{S}$, representing user $u$'s historical interactions, is a sequence of items $\mathcal{S}^u = (i^u_1, i^u_2, \cdots, i^u_{|\mathcal{S}^u|})$ for $u \in \mathcal{U}$. 
% The $k$-th interaction of user $u$, denoted as $i^u_k$, corresponds to the index of the interacted item in the item set $\mathcal{I}$. 
Moreover, each item $i \in \mathcal{I}$ is associated with title and description text $(t^i, d^i) \in \mathcal{T}$.

% where $\mathcal{S}^{u} \in \mathcal{S}$ represents user $u$'s historical interaction $\mathcal{S}^{u} = (i^{u}_{1}, i^{u}_{2}, \cdots, i^{u}_{|\mathcal{S}^{u}|})$ for $u \in \mathcal{U}$. The user $u$'s $k$-th interaction $i^{u}_{k}$ is the index of the interact item in the item set $\mathcal{I}$. Each item $i\in\mathcal{I}$ has item title information and description information $(t_i, d_i) \in \mathcal{T}$, respectively. 

\smallskip
\noindent \textbf{Task: Sequential Recommendation.} The goal of sequential recommendation is to predict the next item to be interacted with by a user based on the user's historical interaction sequence. 
Given a set of user historical interaction sequences $\mathcal{S} = \left\{\ \mathcal{S}^{1}, \mathcal{S}^{2}, \cdots, \mathcal{S}^{|\mathcal{U}|}\right\}$, where $\mathcal{S}^{u}$ denotes the sequence of user $u$, the subset $\mathcal{S}^{u}_{1:k} \subseteq \mathcal{S}^{u}$ represents the sequence of user $u$ from the first to the $k$-th item denoted as $\mathcal{S}^{u}_{1:k} = (i^{u}_{1}, i^{u}_{2}, \cdots, i^{u}_{k})$. 
% For each user's historical interaction $\mathcal{S}^{u}$, let the subset of $\mathcal{S}^{u}_{1:k} \subseteq \mathcal{S}^{u}$ denotes, the user historical interaction from the first item to k-th item, $\mathcal{S}^{u}_{1:k} = (i^{u}_{1}, i^{u}_{2}, \cdots, i^{u}_{k})$.
Given an item embedding matrix $\mathbf{E} \in \mathbb{R}^{|I|\times d}$,
% \footnote{Note that modality-aware recommenders that utilize pre-trained encoders such as BERT and dataset $\mathcal{T}$, construct the item embedding matrix}, 
the embedding matrix of items in $\mathcal{S}^{u}_{1:k}$ is denoted by $\mathbf{E}^{u}_{1:k} = (\textbf{E}_{i_1^u}, \textbf{E}_{i_2^u}, ..., \textbf{E}_{i_k^u}) \in \mathbb{R}^{k \times d}$, where $\textbf{E}_{i_j^u}$ denotes the $i_j^u$-th row of $\mathbf{E}$. This sequence embedding matrix is fed into a collaborative filtering recommender (e.g., SASRec~\cite{kang2018self}) to learn and predict the next item in the user behavior sequence $\mathcal{S}^{u}_{1:k}$ as follows:
\begin{equation}
\small
    \underset{\Theta}{\max}\prod_{u\in\mathcal{U}}\prod_{k=1}^{|\mathcal{S}^{u}|-1}p(i^{u}_{k+1}|\mathcal{S}^{u}_{1:k};\Theta)
    \label{Eq sequential recommendation}
\end{equation}
where $p(i^{u}_{k+1}|\mathcal{S}^{u}_{1:k};\Theta)$ represents the probability of the $(k+1)$-th interaction of user $u$ conditioned on the user's historical interaction sequence $\mathcal{S}^{u}_{1:k}$, and $\Theta$ denotes the set of learnable parameters of the collaborative filtering recommender (CF-RecSys). By optimizing $\Theta$ to maximize Equation \ref{Eq sequential recommendation}, 
% \textcolor{red}{the model can obtain the user representation for user $u$, enabling it to recommend suitable items for users. (CY: This is unclear.)}
{the model can obtain the probability of the next items for user $u$, over all possible items.}

It is important to note that although we mainly focus on the sequential recommendation task in this work,~\proposed~can also be readily applied to non-sequential recommendation tasks by simply replacing the backbone CF-RecSys, e.g., from SASRec~\cite{kang2018self} (sequential) to NCF~\cite{he2017neural} (non-sequential), which will be demonstrated in the experiments {(Section~\ref{exp: model_agnostic})}.

% \subsection{Prompt Design}
% \label{Sec Prompt Engineering}
% \textcolor{red}{
% Even though LLMs demonstrate strong proficiency in general natural language understanding and generation, LLMs still struggle with newly emerged tasks in which LLMs are not trained in the pre-train stage. Hence, designing a proper prompt 
% prompt engineering help in understanding the capabilities and limitations of LLMs, enabling them to perform complex tasks such as question answering and arithmetic reasoning. 
% Because in this study, we aim to utilize an LLM to recommend items to users from a set of candidate items, we suggest our prompt design example in Figure \ref{fig: prompt}.
% }

\section{Proposed Method:~\proposed}
In this section, we propose ~\proposed, a novel LLM-based recommender framework that aligns a frozen pre-trained collaborative filtering recommender (CF-RecSys) with a frozen LLM aiming to enhance the recommendation performance not only in the cold scenario but also in the warm scenario. To bridge the modality gap,~\proposed~aligns collaborative knowledge of the CF-RecSys with the token space of the LLM.
Our approach involves two pre-training stages: (1) Aligning collaborative and textual knowledge with a frozen CF-RecSys (Section \ref{Sec phase1}), and (2) Recommendation stage with a frozen LLM (Section \ref{Sec phase2}) in which the joint collaborative and textual knowledge is projected onto the LLM.

\begin{figure*}[t]
    \centering
    \includegraphics[width=0.89\linewidth]{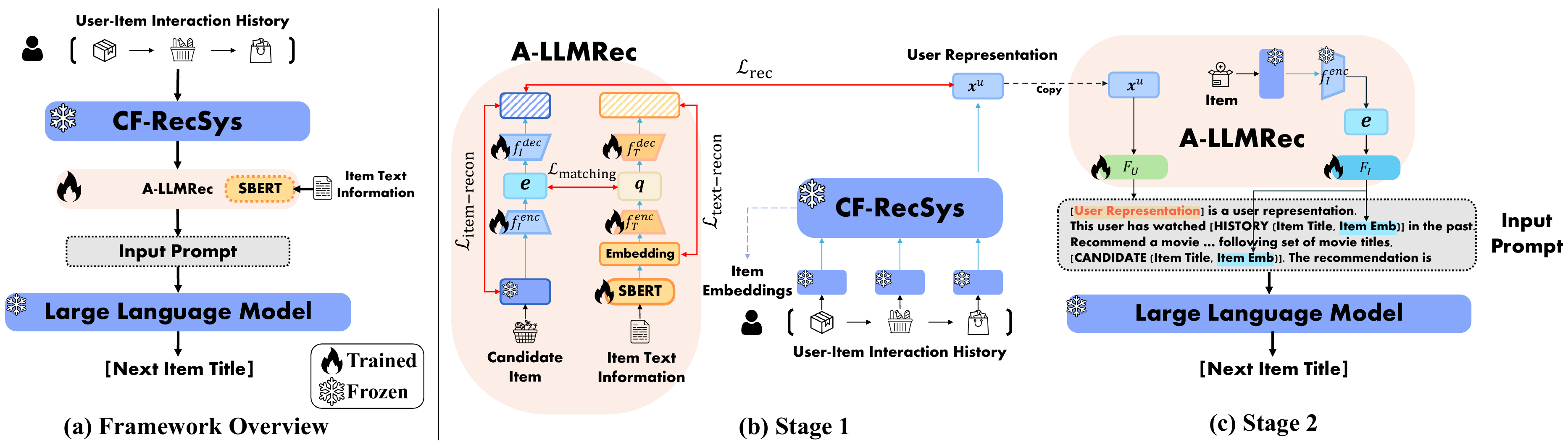}
    \vspace{-2ex}
    \caption{(a) is the overview of~\proposed. (b) and (c) are the detailed architecture of Stage 1 and Stage 2, respectively.}
    \label{fig: framework}
    \vspace{-2ex}
\end{figure*}

% \begin{figure}[t]
%     \centering
%     \includegraphics[width=0.9\linewidth]{imgs/framework_stage1.pdf}
%     \vspace{-2ex}
%     \caption{Model framework stage 1. \textcolor{red}{(CY: Shouldn't SBERT be also frozen? Be clear about which ones are frozen and which ones are trainable. Put snow and fire icon as in TERACON along with a legend. )}}
%     \label{fig: framework stage 1}
%     \vspace{-2ex}
% \end{figure}

% \begin{figure}[t]
%     \centering
%     \includegraphics[width=0.9\linewidth]{imgs/framework_stage2.pdf}
%     \vspace{-2ex}
%     \caption{Model framework stage 2.}
%     \label{fig: framework stage 2}
%     \vspace{-2ex}
% \end{figure}

\subsection{Alignment between Collaborative and Textual Knowledge (Stage-1)}
\label{Sec phase1}
In this section, we introduce how to align the item embeddings from a frozen CF-RecSys with their associated text information to capture both collaborative and textual knowledge.
We employ a pre-trained Sentence-BERT (SBERT) \cite{reimers2019sentence} model, {which is fine-tuned during training}, to extract text embeddings from textual information associated with items\footnote{Although using a larger language model, such as OPT \cite{zhang2022opt} and LLaMA \cite{touvron2023llama}, would further enhance the quality of the text embeddings, we adopt SBERT for efficiency.}.
% Due to the high resource and time requirements for extracting textual information from LLMs, we utilize pre-trained SBERT \cite{reimers2019sentence} which is trained on the Natural Question dataset, with a token size of 768, for extracting textual information.
Then, we introduce two encoders, i.e., item encoder $f_{I}^{enc}$ and text encoder $f_{T}^{enc}$, each containing a 1-layer Multi-Layer Perceptron (MLP), to align the item embeddings from a frozen CF-RecSys with the text embeddings from SBERT.
% To align the embeddings of items and textual data derived from SBERT and a frozen recommender system (RecSys), we design two auto-encoders: 
% one linked to the frozen RecSys and the other connected to SBERT, facilitating latent space alignment. Each autoencoder comprises a single-layer Multi-Layer Perceptron (MLP) for both the encoder and decoder components.
% To align item embeddings and text embeddings from SBERT and frozen RecSys, we construct two auto-encoders: one connected to the frozen RecSys and the other to SBERT, performing latent space matching. Each auto-encoder consists of a 1-layer MLP for both the encoder and decoder.
Given an item $i$, the item encoder $f_{I}^{enc}:\mathbb{R}^d\rightarrow \mathbb{R}^{d'}$ encodes an item embedding $\mathbf{E}_i\in\mathbb{R}^d$ into a latent item embedding $\mathbf{e}_i\in\mathbb{R}^{d'}$, i.e., $\mathbf{e}_i=f_{I}^{enc}(\mathbf{E}_i)$, while the text encoder $f_{T}^{enc}:\mathbb{R}^{768}\rightarrow\mathbb{R}^{d'}$ encodes a text embedding $\mathbf{Q}_i\in\mathbb{R}^{768}$ from SBERT, whose output dimension size is 768, into a latent text embedding $\mathbf{q}_i\in\mathbb{R}^{d'}$, i.e., $\mathbf{q}_i=f_{T}^{enc}(\mathbf{Q}_i)$.
% Likewise, the second autoencoder encodes a text embedding $\mathbf{Q}_i$ through $f_{T}^{enc}$ into a latent text embedding $\mathbf{q}_i$, i.e., $\mathbf{q}_i=f_{T}^{enc}(\mathbf{Q}_i)$,
% and
% Denote $f_{I}^{enc}, f_{I}^{dec}$ are the encoder and decoder of item embedding, i.e., $\mathbf{W}_{I}^{enc} \in \mathbb{R}^{d \times d'}, \mathbf{W}_{I}^{dec} \in \mathbb{R}^{d' \times d}$. For the SBERT, we construct $f_{T}^{enc}, f_{T}^{dec}$ which are the encoder and decoder of text embedding, i.e., $\mathbf{W}_{T}^{enc} \in \mathbb{R}^{768 \times d'}, \mathbf{W}_{T}^{dec} \in \mathbb{R}^{d' \times 768}$. 
Then, we perform latent space matching between item embeddings and text embeddings as follows:
% For the time complexity and memory efficiency, we 
\begin{equation}
\small
\begin{split}
    \mathcal{L}_\text{matching} &= \underset{\mathcal{S}^u\in\mathcal{S}}{\mathbb{E}}\left[\underset{i\in \mathcal{S}^u}{\mathbb{E}}\left[MSE(\mathbf{e}_i, \mathbf{q}_i)\right]\right] \\
    &= \underset{\mathcal{S}^u\in\mathcal{S}}{\mathbb{E}}\left[\underset{i\in \mathcal{S}^u}{\mathbb{E}}\left[MSE(f_{I}^{enc}(\mathbf{E}_i), f_{T}^{enc}(\mathbf{Q}_i))\right]\right]
\end{split}
    \label{Eq matching loss}
\end{equation}
% \begin{equation}
% \begin{split}
%     \mathcal{L}_\text{matching} &= \underset{\mathcal{S}^u\in\mathcal{S}}{\mathbb{E}}\left[\underset{i =  \mathcal{S}^u_{|\mathcal{S}^u|}, \mathcal{S}^{u,-}_{|\mathcal{S}^u|}}{\mathbb{E}}\left[MSE(\mathbf{e}_i, \mathbf{q}_i)\right]\right] \\
%     &= \underset{\mathcal{S}^u\in\mathcal{S}}{\mathbb{E}}\left[\underset{i = \mathcal{S}^u_{|\mathcal{S}^u|}, \mathcal{S}^{u,-}_{|\mathcal{S}^u|}}{\mathbb{E}}\left[MSE(f_{I}^{enc}(\mathbf{E}_i), f_{T}^{enc}(\mathbf{Q}_i))\right]\right]
% \end{split}
%     \label{Eq matching loss}
% \end{equation}
% \begin{equation}
%     \mathcal{L}_\text{match} = MSE(\mathbf{e}_i, \mathbf{q}_i) = MSE(f_{I}^{enc}(\mathbf{E}_i), f_{T}^{enc}(\mathbf{Q}_i))
%     \label{Eq matching loss}
% \end{equation}
where $\mathbf{Q}_i = \textit{SBERT}(``Title: t^i, Description: d^{i}")$ denotes the encoded representation of item text (i.e., item title and description) by SBERT, 
% \textcolor{blue}{$\mathcal{S}^u_{|\mathcal{S}^u|}$ is user $u$'s last interacted item, while $\mathcal{S}^{u,-}_{|\mathcal{S}^u|}$ corresponds to a negative item,}
% \textcolor{blue}{$\mathcal{Y}^u = \left\{i_l^{u,pos},i_l^{u,neg}\right\}$ is set of a positive and negative item of user $u$, respectively, }
and $MSE$ is the mean squared error loss. 
That is, we match the item embeddings from a frozen CF-RecSys and the text embeddings from SBERT in the latent space of the encoders, so as to align the semantics of items and their associated texts for later use in the LLM. 
% It is important to note that for each user $u$ we only considered the last item in $\mathcal{S}^u$ to minimize Equation~\ref{Eq matching loss} instead of all items in $\mathcal{S}^u$ for efficiency in training. However, considering more items further enhances the recommendation performance, which will be shown in Section~\ref{sec:autoregressive}.

% We define $\mathbf{h}_i=\sigma(f_{I}^{enc}(\mathbf{e}_i))$ as the joint collaborative/text embedding (shortly joint embedding) of item $i$, where $\mathbf{e}_i \in \mathbf{E}$, which contains aligned information from both user-item interaction (i.e., collaborative knowledge) and text. These joint embeddings introduce the collaborative and textual information to LLMs in Section \ref{Sec phase2}.

% \smallskip
% \noindent\textbf{Avoiding Over-smoothed Representation. }
\subsubsection{Avoiding Over-smoothed Representation}
\label{sec: reconstruction}
On the other hand, simply optimizing the latent space matching loss defined in Equation~\ref{Eq matching loss} would result in over-smoothed representations, i.e., the encoders would be trained to produce similar outputs (i.e., $\mathbf{e}_i\approx \mathbf{q}_i$) to minimize $\mathcal{L}_\text{matching}$. In an extreme case, the output of the encoders would be collapsed to a trivial representation by assigning their weights to all zeros.
% decodes it back to $\mathbf{\tilde{E}}_i$ through $ f_{I}^{dec}$, i.e., $\mathbf{\tilde{E}}_i=f_{I}^{dec}(\mathbf{e}_i)$. 
% decodes it back to $\mathbf{\tilde{Q}}_i$ through $f_{T}^{dec}$, i.e., $\mathbf{\tilde{Q}}_i=f_{T}^{dec}(\mathbf{q}_i)$.
% entails a trivial solution, because a collapsed output of the encoders would make $\mathcal{L}_\text{match}$ to be equal to zero. 
% While the encoders $f^{enc}_I$ and $f^{enc}_T$ match the information between item embeddings and texts, naively matching this information could lead to the over-smoothing of all joint embeddings. 
% In an extreme case, the weight matrices $\mathbf{W}_{I}^{enc}$ and $\mathbf{W}_{T}^{enc}$ could become zero-value matrices. 
Hence, to prevent this issue and preserve the original information of the item and its associated text embedding, we add a decoder to each of the encoders and introduce reconstruction losses as follows:
\begin{equation}
\small
    \mathcal{L}_\text{item-recon} = \underset{\mathcal{S}^u\in\mathcal{S}}{\mathbb{E}}\left[\underset{i \in \mathcal{S}^u}{\mathbb{E}}\left[MSE(\mathbf{E}_i, f_{I}^{dec}(f_{I}^{enc}((\mathbf{E}_i)))\right]\right]
    \label{Eq item reconstruction loss}
\end{equation}
% \begin{equation}
%     \mathcal{L}_\text{item-recon} = \underset{\mathcal{S}^u\in\mathcal{S}}{\mathbb{E}}\left[\underset{i =  \mathcal{S}^u_{|\mathcal{S}^u|}, \mathcal{S}^{u,-}_{|\mathcal{S}^u|}}{\mathbb{E}}\left[MSE(\mathbf{E}_i, f_{I}^{dec}(f_{I}^{enc}((\mathbf{E}_i)))\right]\right]
%     \label{Eq item reconstruction loss}
% \end{equation}
% \begin{equation}
%     \mathcal{L}_\text{item-recon} = \underset{\mathcal{S}^u\in\mathcal{S}}{\mathbb{E}}\left[MSE(\mathbf{E}_i, f_{I}^{dec}(f_{I}^{enc}((\mathbf{E}_i)))\right]
%     \label{Eq item reconstruction loss}
% \end{equation}
\begin{equation}
\small
    \mathcal{L}_\text{text-recon} = \underset{\mathcal{S}^u\in\mathcal{S}}{\mathbb{E}}\left[\underset{i\in\mathcal{S}^u}{\mathbb{E}}\left[MSE(\mathbf{Q}_i, f_{T}^{dec}(f_{T}^{enc}((\mathbf{Q}_i)))\right]\right]
    \label{Eq Text reconstruction loss}
\end{equation}
% \begin{equation}
%     \mathcal{L}_\text{text-recon} = \underset{\mathcal{S}^u\in\mathcal{S}}{\mathbb{E}}\left[\underset{i =  \mathcal{S}^u_{|\mathcal{S}^u|}, \mathcal{S}^{u,-}_{|\mathcal{S}^u|}}{\mathbb{E}}\left[MSE(\mathbf{Q}_i, f_{T}^{dec}(f_{T}^{enc}((\mathbf{Q}_i)))\right]\right]
%     \label{Eq Text reconstruction loss}
% \end{equation}
% \begin{equation}
%     \mathcal{L}_\text{text-recon} = MSE(\mathbf{Q}_i, f_{T}^{dec}(f_{T}^{enc}((\mathbf{Q}_i)))
%     \label{Eq Text reconstruction loss}
% \end{equation}
where $f_{I}^{dec}$ and $f_{T}^{dec}$ are the decoders added to the encoders $f_{I}^{enc}$ and $f_{T}^{enc}$, respectively. In Section~\ref{exp: ablation for losses in stage1}, we empirically demonstrate the benefit of introducing the reconstruction losses. 
% Note that for efficiency in training, as in Equation~\ref{Eq matching loss}, we only considered the last item in $\mathcal{S}^u$ for each user $u$ to minimize Equations~\ref{Eq item reconstruction loss} and~\ref{Eq Text reconstruction loss}. 
% In {Section~\ref{sec:autoregressive}}, we empirically show that considering more items further enhances the recommendation performance.

% In experiment \ref{exp: ablation for losses in stage1}, we empirically proved the necessity of each loss.
% \smallskip
% \noindent\textbf{Recommendation Loss. }
\subsubsection{Recommendation Loss}
Besides aligning the collaborative knowledge from the user-item interactions with the textual knowledge from the associated text information, we introduce a recommendation loss to explicitly incorporate the collaborative knowledge, while informing the model about the recommendation task. 
Specifically, the recommendation loss is defined as follows~\cite{kang2018self}:
% \begin{equation}
% \small
% \begin{split}
%     \mathcal{L}_\text{rec} = -\sum_{\mathcal{S}^{u} \in \mathcal{S}}
%     \sum_{k =1}^{|\mathcal{S}^u|-1}
%     \big[ log(s(\mathbf{x}^{u}_{k},f^{dec}_{I}(f^{enc}_I(\mathbf{E}_{i^{u}_{k+1}})))) \\ + log(s(\mathbf{x}^{u}_{k},f^{dec}_{I}(f^{enc}_I(\mathbf{E}_{i^{u,neg}_{k+1}}))))\big]
% \end{split}
% \label{Eq Recommendation loss}
% \end{equation}
\begin{equation}
\small
\begin{split}
    \mathcal{L}_\text{rec} = -\sum_{\mathcal{S}^{u} \in \mathcal{S}}
    \big[ log(\sigma(s(\mathbf{x}^{u}_{|\mathcal{S}^u|-1},f^{dec}_{I}(f^{enc}_I(\mathbf{E}_{i^{u}_{|\mathcal{S}^u|}}))))) \\ + log(1-\sigma(s(\mathbf{x}^{u}_{|\mathcal{S}^u|-1},f^{dec}_{I}(f^{enc}_I(\mathbf{E}_{i^{u,-}_{|\mathcal{S}^u|}})))))\big]
\end{split}
\label{Eq Recommendation loss}
\end{equation}
% where $s(\mathbf{a},\mathbf{b})$ is a dot product between $\mathbf{a}$ and $\mathbf{b}$, $\mathbf{x}^{u}_{k} = \textsc{RecSys}(\mathcal{S}^{u}_{1:k})\in\mathbb{R}^{d}$ is the user representation extracted from the collaborative recommender system, i.e., \textsc{RecSys}, obtained after the user $u$ has interacted with the $k$-th item $i^u_{k}$ in the sequence $\mathcal{S}^{u}_{1:k}$, and  $\mathbf{E}_{i^{u,neg}_{k+1}}\in\mathbb{R}^{d}$ is the embedding of a negative item of $i^{u}_{k+1}$, i.e., $i^{u,neg}_{k+1}$.
{where $\mathbf{x}^{u}_{|\mathcal{S}^u|-1} = \textsc{CF-RecSys}(\mathcal{S}^{u}_{1:|\mathcal{S}^u|-1})\in\mathbb{R}^{d}$ is the user representation extracted from the collaborative filtering recommender system, i.e., \textsc{CF-RecSys}, obtained after the user $u$ has interacted with the last item in the sequence $\mathcal{S}^{u}_{1:|\mathcal{S}^u|-1}$, and $\mathbf{E}_{i^{u,-}_{|\mathcal{S}^u|}}\in\mathbb{R}^{d}$ is the embedding of a negative item of $i^{u}_{|\mathcal{S}^u|}$, i.e., $i^{u,-}_{|\mathcal{S}^u|}$, and $s(\mathbf{a},\mathbf{b})$ is a dot product between $\mathbf{a}$ and $\mathbf{b}$.}

% Let $l$ denote the length of user sequence
% Let the user's interaction sequence length of training data be $l$, denote user $u$'s embedding at before the last interaction be $\mathbf{x}^{u}_{l-1} = \textit{RecSys}(\mathcal{S}^{u}_{1:l-1})$, where $\textit{RecSys}$ is the RecSys, the recommendation loss defined as:
% \begin{equation}
% \begin{split}
%     \mathcal{L}_\text{rec} = -\sum_{\mathcal{S}^{u} \in \mathcal{S}} \big[ log(\sigma(\mathbf{x}^{u}_{l-1}(f^{dec}_{T}(\mathbf{h}^{u,pos}_{l}))^T)) \\ + log(1-\sigma(\mathbf{x}^{u}_{l-1}(f^{dec}_{T}(\mathbf{h}^{u,neg}_{l}))^T)\big]
% \end{split}
% \label{Eq Recommendation loss}
% \end{equation}
% where $\mathbf{h}^{u,pos}_{l}$ and $\mathbf{h}^{u,neg}_{l}$ are the joint embedding from the pair of positive and negative item (i.e., $i^{u,pos}_{l}$ and $i^{u,neg}_{l}$) that user $u$ will interact at $l$. 
% Through the collaborative knowledge from $\mathbf{x}$, the $\mathbf{h}$ implicitly learns the collaborative information, which will be provided to LLM.

% \smallskip
% \noindent\textbf{Final Loss of Stage-1. }
\subsubsection{Final Loss of Stage-1}
Finally, the final objective of Stage-1, i.e., $\mathcal{L}_\text{stage-1}$, is the sum of the matching loss defined in Equation \ref{Eq matching loss}, reconstruction losses defined in Equation \ref{Eq item reconstruction loss} and \ref{Eq Text reconstruction loss}, and recommendation loss in Equation \ref{Eq Recommendation loss}:
\begin{equation}
\small
    \mathcal{L}_\text{stage-1}= \mathcal{L}_\text{matching} + \alpha \mathcal{L}_\text{item-recon} + \beta\mathcal{L}_\text{text-recon} + \mathcal{L}_\text{rec}
    \label{Eq overall loss}
\end{equation}
where $\alpha$ and $\beta$ are the coefficients that control the importance of each term. Note that for efficiency in training, we only considered the last item in $\mathcal{S}^u$ for each user $u$ to minimize $\mathcal{L}_\text{stage-1}$. However, considering all items in the sequence further enhances the recommendation performance, which will be shown in Section~\ref{sec:autoregressive}.
% textcolor{red}{(CY: Each loss is only regarding item $i$, which is misleading.)}
% , and $\mathcal{L}$ is optimized using the Adam optimizer. 

%%%%%%%%%% This should go to implementation detail
% For training efficiency, we optimize Equation \ref{Eq overall loss} using only the last interaction information on the user, specifically the item embeddings $\mathbf{e}^{u,pos}_{l}$ and $\mathbf{e}^{u,neg}_{l}$, and the user embedding $\mathbf{x}^{u}_{l-1}$. More precisely, the model is not trained auto-regressive manner. We use only the sequence of user interactions just before the last interaction where sequence length $l-1$, and use the last item as a positive/negative item at $l$.

% \smallskip
% \noindent\textbf{Joint Collaborative-Text Embedding. }
\subsubsection{Joint Collaborative-Text Embedding}
\label{sec: joint embedding}
Having trained the autoencoder based on Equation~\ref{Eq overall loss}, we consider $\mathbf{e}_i=f_{I}^{enc}(\mathbf{E}_i)$ as the joint collaborative-text embedding (shortly joint embedding) of item $i$, which will be passed to the LLM as input. The joint embedding introduces the collaborative and textual knowledge to LLMs, which will be described in Section~\ref{Sec phase2}.

It is important to note that when encountering new items that have not been seen during the training of the collaborative filtering recommender, we can instead rely on the text encoder $f_{T}^{enc}$ to extract the joint collaborative-text embedding, i.e., $\mathbf{q}_i=f_{T}^{enc}(\mathbf{Q}_i)$. Since the two encoders $f_{I}^{enc}$ and $f_{T}^{enc}$ are jointly trained to match their latent spaces, we expect the joint embedding $\mathbf{q}_i$ to not only capture the textual knowledge but also to implicitly capture the collaborative knowledge. In summary, we use $\mathbf{e}_i=f_{I}^{enc}(\mathbf{E}_i)$ as the joint collaborative-text embedding by default, but we use $\mathbf{q}_i=f_{T}^{enc}(\mathbf{Q}_i)$ when item $i$ lacks interactions, i.e., cold item, few-shot, and {cross-domain scenarios}, which will be demonstrated in the experiments in {Section~\ref{exp: cold/warm item}, 
Section~\ref{exp: few-shot}, and Section~\ref{exp: cross-domain}}, respectively.
% \textcolor{red}{TODO: Add remarks on using the other branch.}

% Item, User Representation
% LLM learning method + sequential recommendation
% dimension
% auto-regressive
% \subsection{Projecting Joint Embedding onto LLM (Stage-2)}
\subsection{Alignment between Joint Collaborative-Text Embedding and LLM (Stage-2)}
\label{Sec phase2}
Recall that in Stage-1 we obtained the joint collaborative-text embeddings by aligning the collaborative knowledge with item textual information. Our goal in Stage-2 is to align these joint embeddings with the token space of the LLM (Section~\ref{sec:LLM1}), and design a prompt that allows the LLM to solve the recommendation task by leveraging the learned collaborative knowledge (Section~\ref{sec:LLM2}). Figure~\ref{fig: framework} shows the overall architecture of Stage-2.
{Note that the component trained in Stage-1, which is also utilized in Stage-2, i.e., $f_I^{enc}$, is frozen in Stage-2.}

% In this section, we 
% To this end, we utilize linear layers to adjust and map both joint embeddings $\mathbf{e}$ and user representations $\mathbf{x}$ onto the LLMs (Section~\ref{sec:LLM1}). 
% Note that the parameters of layers involved in Stage-1 are not updated, since they already match the textual information and collaborative knowledge. 
% updating their parameters is costly and disrupts the alignment.
% Furthermore, we introduce a novel prompting approach that merges collaborative knowledge into the LLM (Section~\ref{sec:LLM2}). This is done by incorporating user representations and joint embeddings into the textual prompts in the token embedding space. 
% Figure \ref{fig: framework stage 2} shows the overall architecture of Stage-2.

\subsubsection{Projecting collaborative knowledge onto the token space of LLM}\label{sec:LLM1}
We first project the user representations $\textbf{x}^u\in\mathbb{R}^{d}$ and the joint collaborative-text embeddings $\textbf{e}_i\in\mathbb{R}^{d'}$ obtained from Stage-1 onto the token space of LLM, i.e., $\mathbb{R}^{d^{\text{token}}}$. By doing so, we allow the LLM to take them as inputs. More precisely, we introduce two 2-layer MLPs, i.e., $F_{U}:\mathbb{R}^{d} \rightarrow \mathbb{R}^{d^{\text{token}}}$ and $F_{I}:\mathbb{R}^{d'} \rightarrow \mathbb{R}^{d^{\text{token}}}$, to project the user representations and the joint collaborative-text embeddings to the token space of LLM, respectively, as follows:
\begin{equation}
\small
    \mathbf{O}_u = F_{U}(\mathbf{x}^u), \,\,\mathbf{O}_i = F_{I}(\mathbf{e}_i)
    \label{User representation projection on LLM}
\end{equation}
% \begin{equation}
%     \mathbf{O}_i = F_{I}(\mathbf{e}_i),
%     \label{Item embedding projection on LLM}
% \end{equation}
where $\mathbf{O}_u\in\mathbb{R}^{d^\text{token}}$ and $\mathbf{O}_i\in\mathbb{R}^{d^\text{token}}$ are the projected embeddings of the representation of user $u$ and the joint collaborative-text embedding of item $i$, and they can now be used as inputs to LLM prompts, which allow the LLM to perform recommendation without any fine-tuning.

% $\mathbf{O}_u\in\mathbb{R}^{d^\text{token}}$ and $\mathbf{O}_i\in\mathbb{R}^{d^\text{token}}$ from Equation \ref{User representation projection on LLM} are used as soft prompts

% converted into soft prompts. These prompts provide recommendation tasks and integrate both collaborative and textual knowledge into the LLM, hence projected embeddings aid in conducting recommendation tasks without fine-tuning on the LLM.

\begin{figure}[t]
    \centering
    \includegraphics[width=0.9\linewidth]{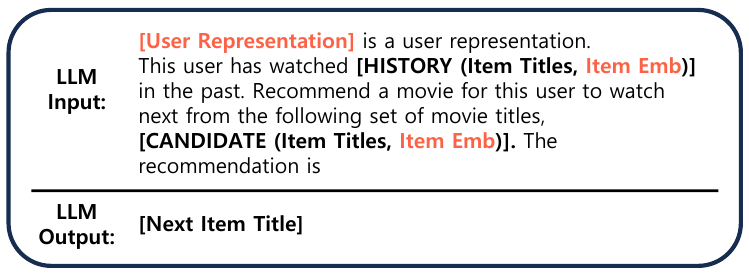}
    \vspace{-2ex}
    \caption{An example prompt of~\proposed~designed for the Amazon Movies dataset. For other datasets, we keep the same format but adjust the verbs and nouns to fit the context (e.g., `watched' $\rightarrow$ `bought', `movie' $\rightarrow$ 'item').}
    \label{fig: prompt}
    \vspace{-2ex}
\end{figure}

\subsubsection{Prompt Design for Integrating Collaborative Knowledge}\label{sec:LLM2}
Prompt engineering helps in understanding the capabilities and limitations of LLMs, enabling them to perform complex tasks such as question answering and arithmetic reasoning~\cite{brown2020language,wei2022chain}. Recent studies on LLM-based recommender systems have shown that carefully crafted prompts enhance the performance of LLMs~\cite{sanner2023large,he2023large,bao2023tallrec}. 
% \textcolor{blue}{These approaches only utilized the LLMs' natural language comprehension and via in-context learning while LLMs are showcasing their remarkable understanding of modality information ~\cite{Li2023BLIP2BL,tsimpoukelli2021multimodal}.} 
However, as existing LLM-based recommender systems focus on cold scenarios with few user-item interactions, their prompts mainly consider ways to incorporate modality information (e.g., item description text), while overlooking the collaborative knowledge.
% \textcolor{red}{(CY: Can you add remark regarding a limitation of existing LLM-based RecSys in terms of prompts? Why makes ours a `novel' approach to prompt design?)}
% Despite their effectiveness in cold scenarios, they fall short of explicitly capturing the collaborative knowledge that is crucial in warm scenarios
% Thus, our main goal in designing prompts is to provide the LLM with a clear recommendation instruction and the collaborative knowledge. 
To this end, we introduce a novel approach to prompt design for LLM-based recommender system, which combines collaborative knowledge with recommendation instructions (See Figure~\ref{fig: prompt}).
% novel prompting approach that merges collaborative knowledge into the LLM.
This is done by directly incorporating user representations $\mathbf{O}_u$ and joint collaborative-text embeddings $\mathbf{O}_i$ into the textual prompts in the token embedding space. 
% Prompt engineering helps in understanding the capabilities and limitations of LLMs, enabling them to perform complex tasks such as question answering and arithmetic reasoning. Also, in the recommendation domain, recent studies have shown that carefully crafted prompts enhance the performance of LLMs. 
% Thus, we aim to provide clear recommendation instruction and collaborative knowledge in prompts. We suggest a new approach to prompt design, which combines collaborative knowledge with recommendation instructions LLM in Figure \ref{fig: prompt}. 
% In the token embedding space of the LLM, we position both the projected user representation, denoted as $\mathbf{O}_u$, and the projected joint embedding of item, $\mathbf{O}_i$. 
In other words, as $\mathbf{O}_u$ and $\mathbf{O}_i$ have been projected into the LLM token space, they can be considered as ordinary tokens used by the LLM and readily incorporated within a prompt. 
To facilitate the understanding of the LLM regarding the given user, which is crucial for personalized recommendation, we place the projected user representation $\mathbf{O}_u$ at the beginning of the prompt to provide the LLM with the information about users, which is analogous to soft prompts~\cite{li-liang-2021-prefix}.
Moreover, we add the projected joint embedding of an item $\mathbf{O}_i$ next to its title.
This structured prompt then serves as an input to the LLM, with the expected output being recommendations tailored to the user. The learning objective of Stage-2 is given as follows:
% \begin{equation}
%     % \underset{\Theta}{\max}\sum_{(x,y)\in\mathcal{Z}}\sum_{t=1}^{|{y}|},
%     % log(P_{\Theta} (y_t |x, y_{<t}))
%     \mathcal{L}_{LLM} = -\sum_{u \in \mathcal{U}}\sum_{k=1}^{|t^{i^u_l,pos}|}log(P^{LLM}_{\theta}(t^{i^u_l,pos}_{k}|\textit{prompt}^{u_{l-1}}, t^{i^u_l,pos}_{<k})
%     \label{LLM loss}
% \end{equation}
% \begin{equation}
%     \mathcal{L}_{LLM} = -\underset{\mathcal{S}^u\in\mathcal{S}}{\mathbb{E}}\left[log(P^{LLM}_{\theta}( M^{\mathcal{S}^u_l}|\textit{prompt}^{S^{u}_{1:l-1}}_{\textbackslash M^{\mathcal{S}^u_l}})\right]
% \end{equation}
% \textcolor{blue}{where $\theta$ represents the learnable parameters of $F_{U}$ and $F_{I}$, $\textit{prompt}^{S^{u}_{1:l-1}}_{\textbackslash M^{\mathcal{S}^u_l}}$ denotes the masked prompt on the output of LLM, where $M^{\mathcal{S}^{u}_l}$ is the next item title for user $u$ with historical interaction is $\mathcal{S}^u_{1:l-1}$ as in Figure \ref{fig: prompt}, and $P^{LLM}$ is the probability of LLM for the mask $M$, given the prompt.}
\vspace{-2ex}
\begin{equation}
\small
    \underset{\theta}{\max}\sum_{\mathcal{S}^u \in \mathcal{S}}\sum_{k=1}^{\mathcal|y^u|}log(P_{\theta, \Theta}(y^u_k|p^u, y^u_{<k}))
    \label{LLM Loss}
\end{equation}
where $\theta$ denotes the learnable parameters of $F_{U}$ and $F_{I}$, $\Theta$ is the frozen parameters of LLM, $p^u$ and $y^u$ are the input prompt and the next item title of user $u$, respectively. $y^u_k$ is the $k$-th token of $y^u$ and $y^u_{<k}$ represents the tokens before $y^u_k$. Note that we only use the last item of each user sequence to train Equation~\ref{LLM Loss} for efficiency.
% i.e., $t_{i^{u}_{|\mathcal{S}^u|}}$, on user $u$'s last interaction sequence, respectively.
% where $\theta$ is the learnable parameters of $F_{U}$ and $F_{I}$, $\textit{prompt}^{u_{l-1}}$ is the prompt construct with user $u$'s interaction history until $l-1$, $t^{i^u_l,pos}$ is item title of user $u$ interacted at time $l$, $t^i_{k}$ is the $k$-th words of title $t^i$, and $t^i_{<k} = \left\{t^i_{1}, t^i_{2}, \cdots, t^i_{k-1}\right\}$.

% \begin{table}[h]
%   \caption{Statistics of the dataset after preprocessing. Avg. Len denotes the average sequence length of users.}
%   \label{tab:dataset}
%     \vspace{-1ex}
%     \resizebox{0.8\linewidth}{!}{

%   \begin{tabular}{ccccc}
%     \toprule
%     Datasets & \#Users & \#Items & \#Interactions. & Avg. Len\\
%     \midrule
%     Movies and TV & 297,498 & 59,944 & 3,409,147 & 11.46\\
%     Video Games & 64,073 & 33,614 & 598,509 & 8.88\\
%     Beauty & 9,930 & 6,141 & 63,953 & 6.44\\
%     Toys & 30,831 & 61,081 & 282,213 & 9.15\\
%   \bottomrule
% \end{tabular}}
% \vspace{-3ex}
% \end{table}

\begin{table*}[t]
\small
    \caption{Overall model performance (Hit@1) over various datasets. The best performance is denoted in bold.}
    \vspace{-2ex}
    \label{tab: main table}
        \vspace{-1ex}
        \resizebox{0.8\linewidth}{!}{
    \begin{tabular}{c|cccc|ccc|cccc}
    \toprule
    \multirow{2}{*}{} & \multicolumn{4}{c|}{Collaborative filtering} & \multicolumn{3}{c|}{Modality-aware} & \multicolumn{4}{c}{LLM-based}            \\ \cline{2-12} 
                      & NCF & NextItNet   & GRU4Rec   & SASRec    & MoRec     & CTRL  & RECFORMER    & LLM-Only  & TALLRec   & MLP-LLM   & \proposed          \\ \midrule\midrule
    Movies and TV     & 0.4273 &0.5855     & 0.5215    & 0.6154    & 0.4130    & 0.3467  & 0.4865 & 0.0121    & 0.2345    & 0.5838    & \textbf{0.6237}   \\ \midrule
    Video Games       & 0.3159& 0.4305      & 0.4026    & \textbf{0.5402}       & 0.4894   & 0.2354 & 0.4925  & 0.0168    & 0.4403    & 0.4788  & 0.5282  \\ \midrule
    Beauty           & 0.2957& 0.4231      & 0.4131    & 0.5298    & 0.4997    & 0.3963  & 0.4878 & 0.0120    & 0.5542    & 0.5548    & \textbf{0.5809}   \\ \midrule
    Toys             & 0.1849& 0.1415      & 0.1673    & 0.2359    &  0.1728   & 0.1344  &  0.2871   & 0.0141    &   0.0710   & 0.3225    &  \textbf{0.3336}  \\ \bottomrule

    \end{tabular}}
        \vspace{-2ex}
\end{table*}

% \begin{table*}[h]
% \caption{Cold user and few-shot scenario.}
%   \label{tab: fewshot}
%     \vspace{-2ex}
%   \resizebox{0.8\linewidth}{!}{
% \begin{tabular}{c|ccc|ccc|ccc}
% \toprule
% \multirow{2}{*}{} & \multicolumn{3}{c|}{Movies and TV}                                          & \multicolumn{3}{c|}{Video Games}                                            & \multicolumn{3}{c}{Beauty}                                                  \\ \cline{2-10} 
%                   & \multicolumn{1}{c}{Cold User} & \multicolumn{1}{c}{Few (256)} & Few (128) & \multicolumn{1}{c}{Cold User} & \multicolumn{1}{c}{Few (256)} & Few (128) & \multicolumn{1}{c}{Cold User} & \multicolumn{1}{c}{Few (256)} & Few (128) \\ \midrule\midrule
% SASRec            & \multicolumn{1}{c}{0.2589}    & \multicolumn{1}{c}{0.2111}    & 0.1537    & \multicolumn{1}{c}{0.4102}    & \multicolumn{1}{c}{0.1396}     &   0.1089   & \multicolumn{1}{c}{0.4459}    & \multicolumn{1}{c}{0.2243}     &    0.1813   \\ \midrule
% TALLRec           & \multicolumn{1}{c}{}          & \multicolumn{1}{c}{}          &           & \multicolumn{1}{c}{}          & \multicolumn{1}{c}{}          &           & \multicolumn{1}{c}{0.5202}    & \multicolumn{1}{c}{}          &           \\ \midrule
% \proposed          & \multicolumn{1}{c}{0.5272}    & \multicolumn{1}{c}{0.2880}    & 0.2518    & \multicolumn{1}{c}{0.4160}    & \multicolumn{1}{c}{0.2495}   &    0.1608    & \multicolumn{1}{c}{0.5337}    & \multicolumn{1}{c}{}          &           \\ \bottomrule
% \end{tabular}}
%     % \vspace{-1ex}
% \end{table*}

\begin{table}[h]
  \caption{Statistics of the dataset after preprocessing. Avg. Len denotes the average sequence length of users.}
  \label{tab:dataset}
    \resizebox{0.9\linewidth}{!}{

  \begin{tabular}{ccccc}
    \toprule
    Datasets & \#Users & \#Items & \#Interactions. & Avg. Len\\
    \midrule
    Movies and TV & 297,498 & 59,944 & 3,409,147 & 11.46\\
    Video Games & 64,073 & 33,614 & 598,509 & 8.88\\
    Beauty & 9,930 & 6,141 & 63,953 & 6.44\\
    Toys & 30,831 & 61,081 & 282,213 & 9.15\\
  \bottomrule
\end{tabular}}
\end{table}

\section{Experiments}
\subsection{Experimental Setup}
\noindent \textbf{Datasets.} For comprehensive evaluations, we used four datasets from Amazon datasets~\cite{mcauley2015image, 10.1145/2872427.2883037}, i.e., Movies and TV, Video Games, Beauty, and Toys, which consist of comprehensive textual information including "title" and "description." Note that we deliberately selected datasets with varying statistics in terms of number of users and items to conduct an extensive analysis of the models. The statistics for each dataset after preprocessing are presented in Table~\ref{tab:dataset} and we describe details regarding data preprocessing as follows:

\begin{itemize}[leftmargin=0.5cm]
\item \textbf{Movies and TV} To evaluate the models on a large scale, we select about 300K users and 60K items. Following existing studies~\cite{kang2018self,10.1145/3539618.3591932},
we removed users and items with fewer than 5 interactions.
\item \textbf{Video Games} To evaluate the models on moderate-scale data, which is smaller than the Movies and TV dataset, we select about 64K users and 33K items, removing users and items with fewer than 5 interactions, as in the Movies and TV dataset.
\item \textbf{Beauty} To compose a small and cold dataset, we select about 9K users and 6K items, removing users and items with fewer than 4 interactions. To retain some information from user-item feedback, we categorized user ratings by treating items above 3 as positive and all others including non-interacted items as negative.
\item \textbf{Toys} For the evaluation of the models where the number of items is larger than number of users, unlike other datasets, we select about 3K users and 6K items, with the number of items being twice as large as the number of users, and remove users and items with fewer than 4 interactions. Similar to the Beauty dataset, to preserve some information from user-item feedback, we categorize positive and negative items with the criterion of rating 3.
\end{itemize}

\smallskip
\noindent \textbf{Baselines.}
We compare~\proposed~with the following baselines that can be categorized into three types: collaborative filtering recommender systems (NCF~\cite{he2017neural}, NextItNet~\cite{yuan2019simple}, GRU4Rec~\cite{hidasi2015session} and SASRec~\cite{kang2018self}), modality-aware recommender systems (MoRec~\cite{10.1145/3539618.3591932}, CTRL~\cite{li2023ctrl}, and RECFORMER~\cite{10.1145/3580305.3599519}), and LLM-based recommender systems (LLM-Only, TALLRec~\cite{bao2023tallrec} and MLP-LLM). For more detail regarding the baselines, please refer to Appendix \ref{append: baselines}

\begin{table}[]
\caption{Hyperparameter specifications of ~\proposed}
\label{tab: hyper-parameter}
\resizebox{1.0\linewidth}{!}{
\begin{tabular}{c|c|c|c|c|c|c}
\hline
\multirow{2}{*}{} & Learning rate & Learning rate & embedding dim & embedding dim & \multirow{2}{*}{alpha} & \multirow{2}{*}{beta} \\
                  & stage 1       & stage 2       & (CF-RecSys) $d$             & ($f_I^{enc},f_T^{enc}$) $d'$           &                        &                       \\ \hline
Movies and TV     & 0.0001        & 0.0001        & 50            & 128           & 0.5                    & 0.5                   \\ \hline
Video Games       & 0.0001        & 0.0001        & 50            & 128           & 0.5                    & 0.5                   \\ \hline
Beauty            & 0.0001        & 0.0001        & 50            & 128           & 0.5                    & 0.2                   \\ \hline
Toys              & 0.0001        & 0.0001        & 50            & 128           & 0.5                    & 0.2                    \\ \hline
\end{tabular}}
\end{table}

\smallskip
\noindent \textbf{Evaluation Setting.}
We divide user sequences into training, validation, and test sets. For each user sequence, the most recently interacted item, denoted as $i^u_{|\mathcal{S}^u|}$, is used as the test set, while the second most recent user interaction item, $i^u_{|\mathcal{S}^u|-1}$, is used as the validation set. The remaining sequence of items is used as the training set. 
% During the inference stage with the validation set and test set, we feed the model the entire sequence except for the items in the validation set and test set, respectively.
To evaluate the performance of sequential recommendation models, we add 19 randomly selected non-interacted items to the test set, so that the test set of each user contains 1 positive item and 19 negative items. For quantitative comparison, we employ a widely used metric, Hit Ratio at 1 (Hit@1) for all experiments.

\smallskip
\noindent \textbf{Implementation Details.}
Although \proposed~is model-agnostic, in this work, we adopt OPT-6.7B~\cite{zhang2022opt} as the backbone LLM and SASRec~\cite{kang2018self} as the pre-trained CF-RecSys. For fair comparisons, we also used OPT-6.7B as the backbone LLM for other LLM-based models (i.e., LLM-Only, TALLRec~\cite{bao2023tallrec} and MLP-LLM). Moreover, we use SASRec as the CF-RecSys in other modality-aware models (i.e., MoRec~\cite{10.1145/3539618.3591932} and CTRL~\cite{li2023ctrl}), and fix the dimension of item and model embeddings to 50 for all the methods and datasets.
For RECFORMER~\cite{10.1145/3580305.3599519}, we follow the paper and employ Longformer~\cite{beltagy2020longformer} as the backbone network.
% \textcolor{blue}{Also, we followed the details in paper \cite{10.1145/3580305.3599519} for implementing RECFORMER.}
We set the batch size to 128 for all collaborative filtering-based and modality-aware models. Moreover, the batch size is set to 32 for Stage-1 of \proposed, and 4 for MLP-LLM, TALLRec, and Stage-2 of \proposed.
% Moreover, we trained the TALLRec using a batch size of 4 which is smaller than the original paper the reason will be comprehensively addressed in the following paragraph.
We trained Stage-1 of \proposed~for 10 epochs, and Stage-2 of~\proposed~for 5 epochs, and TALLRec is trained for a maximum of 5 epochs.
We use the Adam optimizer to train the models in all datasets. For hyperparameters, we tune the model in certain ranges as follows: learning rate $\eta_1, \eta_2$ in $\left\{0.01,0.001, 0.0005, 0.0001\right\}$ for the training stage each, coefficient $\alpha, \beta$ in $\left\{0.1, 0.2, 0.5, 0.75, 1.0\right\}$ for each, we report the best-performing hyper-parameters for each dataset in Table ~\ref{tab: hyper-parameter}. 
% We report the best-performing hyper-parameters for each dataset in Table ~\ref{tab: hyper-parameter} of Appendix~\ref{append: hyper}. 
We use four NVIDIA GeForce A6000 48GB for the Movies and TV dataset to train LLM-based models, and one NVIDIA GeForce A6000 48GB for other datasets including LLM-based and other models.

\begin{table}[t]
\small
\caption{Results (Hit@1) on cold/warm item scenario. \proposed~(SBERT) is a variant of \proposed~that uses $\mathbf{q}$ instead of $\mathbf{e}$ for inference.}
  \label{tab: cold/warm}
    \vspace{-2ex}
\resizebox{0.83\linewidth}{!}{
\begin{tabular}{c|cc|cc|cc}
\toprule
\multirow{2}{*}{} & \multicolumn{2}{c|}{Movies and TV}              & \multicolumn{2}{c|}{Video Games}        & \multicolumn{2}{c}{Beauty}      \\ \cline{2-7} 
                  & \multicolumn{1}{c}{Cold}    & Warm              & \multicolumn{1}{c}{Cold}      & Warm  & \multicolumn{1}{c}{Cold} & Warm \\ \midrule\midrule
SASRec            & \multicolumn{1}{c}{0.2589}  & 0.6787            & \multicolumn{1}{c}{0.1991}            &   0.5764 & \multicolumn{1}{c}{0.1190} & 0.6312   \\ \midrule
MoRec             & \multicolumn{1}{c}{0.2745}  & 0.4395            & \multicolumn{1}{c}{0.2318}            &  0.4977 & \multicolumn{1}{c}{0.2145} & 0.5425   \\ \midrule
CTRL              & \multicolumn{1}{c}{0.1517}  & 0.3840            & \multicolumn{1}{c}{0.2074}            &   0.2513 & \multicolumn{1}{c}{0.1855} & 0.4711  \\ \midrule
RECFORMER              & \multicolumn{1}{c}{0.3796}  &   0.5449   & \multicolumn{1}{c}{0.3039}     &  0.5377  & \multicolumn{1}{c}{0.3387} &  0.5133 \\ \midrule
TALLRec           & \multicolumn{1}{c}{0.2654}  & 0.2987            & \multicolumn{1}{c}{0.3950}            &    0.4897  & \multicolumn{1}{c}{0.5462} & 0.6124\\ \midrule

\proposed          & \multicolumn{1}{c}{0.5714} & \textbf{0.6880}   & \multicolumn{1}{c}{0.4263}   &   \textbf{0.5970}   & \textbf{0.5605} & \multicolumn{1}{c}{\textbf{0.6414}} \\ \midrule
\proposed~(SBERT)         & \multicolumn{1}{c}{\textbf{0.5772}} & 0.6802 & \multicolumn{1}{c}{\textbf{0.4359}} & 0.5792 & \multicolumn{1}{c}{0.5591}  &   0.6405  \\ \bottomrule
\end{tabular}}
    \vspace{-3.5ex}
\end{table}
\subsection{Performance Comparison}
For comprehensive evaluations of~\proposed, we perform evaluations under various scenarios, i.e., general scenario (Sec.~\ref{exp: overall performance}), cold/warm item scenario (Sec.~\ref{exp: cold/warm item}), cold user scenario (Sec.~\ref{exp: cold-user}), few-shot training scenario (Sec.~\ref{exp: few-shot}), cross-domain scenario (Sec.~\ref{exp: cross-domain}).
\subsubsection{Overall Performance.}
\label{exp: overall performance}
% \noindent\textbf{Overall Performance.}
\looseness=-1
The results of the recommendation task on four datasets are given in Table \ref{tab: main table}.
We have the following observations: 
\textbf{1)} \proposed~outperforms other LLM-based recommender systems that do not consider the collaborative knowledge from user-item interactions (i.e., LLM-Only and TALLRec), implying that the collaborative knowledge is crucial for improving the performance of recommendation in general.
% This improvement attributed to the fact that learning collaborative knowledge from user-item interactions can significantly enhance the modeling of complex behavior patterns present in the recommendation dataset. 
\textbf{2)} We observe that MLP-LLM, which replaces the alignment module of~\proposed~with a simple MLP, underperforms~\proposed. This implies that bridging between CF-RecSys and LLM is a challenging problem and that our proposed two-stage alignment module is beneficial.
\textbf{3)} `LLM-Only' performs the worst among the LLM-based models, implying that naively adopting an LLM based on a prompt designed for the recommendation task is not sufficient. Note that the prompt used by `LLM-Only' is exactly the same as the prompt shown in Figure~\ref{fig: prompt} without user representation and item embeddings.
This again demonstrates the importance of incorporating 
collaborative knowledge into the LLM for improving the recommendation performance.
\textbf{4)} While TALLRec fine-tunes the LLM for the recommendation task, it underperforms a collaborative filtering model, SASRec. This highlights that the text information alone may not generate sufficient knowledge for capturing collaborative knowledge effectively even with fine-tuning the LLM. This again demonstrates the superiority of our alignment module.
\textbf{5)} Although the modality-aware models (MoRec and CTRL) use SASRec as the backbone CF-RecSys, they underperform SASRec. Moreover, {RECFORMER struggles to outperform SASRec despite using Longformer for item text attributes, due to the emphasis on textual information in similarity matching between user and item sentences.} This shows that the modality knowledge might hinder the learning of collaborative knowledge, leading to performance degradation.

\subsubsection{Cold/Warm Item Scenarios.}
% \noindent\textbf{Cold/Warm Item Scenarios}
\label{exp: cold/warm item}
This section evaluates the models under cold/warm item scenarios. 
Items are labeled as `warm' if they belong to the top 35\% of interactions, while those in the bottom 35\% are labeled as `cold' items.
% We categorized a dataset into the top 35\% of interactions and the bottom 35\% of interactions. 
After training each model using all the available data in the training set, we separately
evaluate cold and warm items in the test set (Table~\ref{tab: cold/warm}). 
% And measures their performances only regarding the cold/warm items for each scenario. We can conclude the three things via our results in Table \ref{tab: cold/warm}.
% Drawing from the results in Table~\ref{tab: cold/warm}, 
We make the following observations: 
\textbf{1)} ~\proposed~outperforms all other baselines across both scenarios, which demonstrates that our alignment network indeed allows the LLM to understand and utilize the collaborative knowledge. 
\textbf{2)} On the other hand, TALLRec outperforms SASRec only under cold scenario, whereas SASRec outperforms TALLRec only under warm scenario. This demonstrates the importance of capturing both the collaborative knowledge and the text information to excel in both cold/warm scenarios.
\textbf{3)} \proposed~(SBERT) outperforms~\proposed~under the cold item scenario, while \proposed~generally outperforms \proposed~(SBERT) under the warm item scenario. As discussed in Section~\ref{sec: joint embedding}, this implies that the joint collaborative-text embedding obtained from the text encoder given the text information (i.e., $\mathbf{q_i}=f_T^{enc}(\mathbf{Q}_i)$) is more useful than that obtained from the item encoder given the item embedding (i.e., $\mathbf{e_i}=f_I^{enc}(\mathbf{E}_i)$).

\begin{table}[t]
\small
\centering
\caption{Results (Hit@1) on cold user scenario.}
\label{tab: colduser}
\vspace{-2ex}
\resizebox{0.62\linewidth}{!}{
\begin{tabular}{c|c|c|c}
\toprule
         & Movies and TV & Video Games & Beauty \\ \midrule\midrule
SASRec   & 0.2589        & 0.4048      & 0.4459 \\ \midrule
MoRec   & 0.3918        & 0.3572      & 0.4815 \\ \midrule
CTRL   & 0.2273        & 0.1737      & 0.3902 \\ \midrule
RECFORMER   &  0.4481  & 0.3989 & 0.4644  \\ \midrule
TALLRec  & 0.2143        & 0.3895      & 0.5202 \\ \midrule
MLP-LLM  & 0.4909        & 0.3960       & 0.5276  \\ \midrule
\proposed & \textbf{0.5272}  & \textbf{0.4160}  & \textbf{0.5337} \\ \bottomrule
\end{tabular}}
\vspace{-2ex}
\end{table}

\subsubsection{Cold User Scenarios.}
% \noindent\textbf{Cold User/Few-shot Scenarios}
\label{exp: cold-user}
% As shown in Table~\ref{tab: colduser}, we conducted an analysis of model performance, in scenarios where users with a limited interaction history (i.e., cold user) join the platform. 
Besides evaluations under the cold item scenario, we additionally conduct evaluations under the cold user scenario (Table~\ref{tab: colduser}).
% As shown in Table~\ref{tab: colduser}, we conducted an analysis of model performance, in scenarios when the users have a limited interaction history (i.e., cold user).
To simulate the cold user scenario, we sample users who have interacted with exactly three items, where the last item in the sequence serves as the test set. Then, we use the models trained on the entire set of users except for the sampled users to perform inference on the sampled users.
We observe that~\proposed~consistently outperforms other models in the cold user scenario, while SASRec struggles to perform well, especially on a large dataset, i.e., Movies and TV, due to the lack of collaborative knowledge from users. 
Moreover, LLM-based models demonstrate superior performance in handling cold users as text information becomes useful under cold scenarios. 
% \textbf{3)} TALLRec exhibits inferior performance on the Movies and TV dataset, primarily due to its inherently low performance on the original Movies and TV dataset. One of the key advantages of ~\proposed~ is the ability to combine textual information and collaborative knowledge. In the case of cold users with limited interaction data, ~\proposed~ relies more on textual information while still leveraging collaborative knowledge, which contributes to superior performance.

\begin{table}[t]
\caption{Results (Hit@1) on the few-shot training scenario on various datasets ($K$: num. users in the training set). 
% The 256/128 shot means using only 256/128 users for the train dataset. 
% \proposed~(SBERT) is a variant of \proposed~that uses $\mathbf{q}$ instead of $\mathbf{e}$ for inference.
}
\label{tab: fewshot}
\vspace{-2ex}
 \resizebox{0.92\linewidth}{!}{
\begin{tabular}{c||c|c|c|c|c|c}
\toprule
                               & $K$ & SASRec &MoRec& TALLRec & \proposed & \proposed~(SBERT) \\ \midrule\midrule
\multirow{2}{*}{Movies and TV} & 256      & 0.2111& 0.2208 & 0.1846 & 0.2880   &   \textbf{0.2963}    \\ \cline{2-7} 
                               & 128      & 0.1537& 0.1677 & 0.1654  & 0.2518   &   \textbf{0.2722}     \\ \midrule
\multirow{2}{*}{Video Games}   & 256      & 0.1396& 0.1420 &  0.2321  & 0.2495   &   \textbf{0.2607}        \\ \cline{2-7} 
                               & 128      & 0.1089& 0.1157 &  0.1154   & 0.1608   &  \textbf{0.1839}          \\ \midrule
\multirow{2}{*}{Beauty}        & 256      & 0.2243& 0.2937 &  0.3127 &  0.3467 &  \textbf{0.3605}    \\ \cline{2-7} 
                               & 128      & 0.1813& 0.2554 & 0.2762   &  0.3099    &  \textbf{0.3486}         \\ \bottomrule
\end{tabular}}
\vspace{-3ex}
\end{table}

\subsubsection{Few-shot Training Scenario.}
\label{exp: few-shot}
To investigate the impact of unseen/new items on recommendation models, we conduct experiments on a few-shot training scenario where the number of users in the training set is extremely limited to only $K$ users, i.e., $K$-shot (Table~\ref{tab: fewshot}). 
Under this scenario, we expect the models to encounter a large amount of unseen/new items at the inference stage, which would make it hard to provide accurate recommendations.
% Consequently, the models hard to access a sufficient number of items, which leads to the presence of numerous unseen/new items.
% In Table~\ref{tab: fewshot}, we have the following observations: \textbf{1)} The LLM-based models outperform the C-Resys, i.e., SASRec, due to the textual understanding of LLM, which helps extract information from the text of the unseen item. \textbf{2)} ~\proposed~ outperforms other LLM-based models. Despite training with extremely few data,~\proposed~ captures the knowledge about how collaborative filtering works from C-Resys, and combines this knowledge with the textual information of items, leading to exceptional performance in few-shot learning. \textbf{3)} The SBERT branch enhances the utilization of informative item knowledge in few-shot learning scenarios. As mentioned in Section~\ref{sec: joint embedding}, when the items lack interactions,~\proposed~can use $\mathbf{q}$, which is an informative vector constructed from item texts, as the joint embedding instead of $\mathbf{e}_i$, which is randomly initialized for unseen/new items. Thanks to the alignment between collaborative and textual knowledge (Stage-1), as described in Section~\ref{Sec phase1}, $\mathbf{q}$ captures not only rich textual information of items but also the collaborative knowledge, leading to excelling performance, as shown in the Table~\ref{tab: fewshot}.
We have the following observations: \textbf{1)} ~\proposed~ outperforms all other baselines under the few-shot scenario. Despite being trained with extremely small amount of users, \proposed~relies on CF-RecSys to capture the collaborative knowledge, which is combined with the textual knowledge of items, leading to superior performance in few-shot learning. 
\textbf{2)} \proposed~(SBERT) outperforms \proposed, implying again that using the text encoder to extract the joint text-collaborative knowledge is useful when items lack interactions.
% ,~\proposed~can use $\mathbf{q}$, which is an informative vector constructed from item texts, as the joint embedding instead of $\mathbf{e}$, which is randomly initialized for unseen/new items. Thanks to the alignment between collaborative and textual knowledge (Stage-1), as described in Section~\ref{Sec phase1}, $\mathbf{q}$ captures not only rich textual information of items but also the collaborative knowledge, leading to excelling performance, as shown in the Table~\ref{tab: fewshot}.
\textbf{3)} Under the few-shot scenario, LLM-based models outperform the CF-Resys, i.e., SASRec, due to the textual understanding of LLM, which helps extract information from the text of the unseen item, while CF-RecSys suffers from the lack of collaborative knowledge regarding unseen/new items.
\vspace{-1ex}
\begin{table}[h]
\caption{Results (Hit@1) on a cross-domain scenario (i.e., Pre-trained: Movies and TV, Evaluation: Video Games). 
% \proposed~(SBERT) is a variant of \proposed~that uses $\mathbf{q}$ instead of $\mathbf{e}$ for inference.
}
\label{tab: cross-domain}
    \vspace{-2ex}
\resizebox{1.0\linewidth}{!}{
\begin{tabular}{c|c|c|c|c|c|c}
\hline
      & SASRec & MoRec &RECFORMER & TALLRec & \proposed & \proposed~(SBERT) \\ \midrule\midrule
\begin{tabular}[c]{@{}c@{}}Movies and TV\\ $\rightarrow$ Video Games\end{tabular} & 0.0506  & 0.0624   & 0.0847 & 0.0785   & 0.0901                   & \textbf{0.1203  }                         \\ \bottomrule
\end{tabular}}
    \vspace{-2.5ex}
\end{table}

\subsubsection{Cross-domain Scenario.}
\label{exp: cross-domain}
To further investigate the generalization ability of ~\proposed, we evaluate the models on the cross-domain scenario, where the models are evaluated on datasets that have not been used for training (Table~\ref{tab: cross-domain}). 
% Specifically, we set up a particularly challenging case by not performing fine-tuning on cross-domain datasets for all models, to evaluate the generalization ability of each model and prove ~\proposed~ superiority.
Specifically, we pre-train the models on the Movies and TV dataset and perform evaluations on the Video Games dataset. We have the following observations:
\textbf{1)} \proposed~outperforms all the baselines in the cross-domain scenario, and \proposed~(SBERT) particularly performs well. This is again attributed to the text encoder that becomes useful when collaborative information is lacking.
\textbf{2)} SASRec underperforms modality-aware models and LLM-based models, indicating that using textual knowledge is crucial for the cross-domain scenario due to the lack of collaborative information. 
% which proves simply replacing the joint embedding $\mathbf{e}$ to $\mathbf{q}$ can effectively supplement the lack of collaborative knowledge in cross-domain items as described in Section ~\ref{sec: joint embedding}, which highlight ~\proposed's powerful generalization ability in scenarios where collaborative knowledge is scarce, as cross-domain.
% \textbf{2)} CF-RecSys (SASRec) struggled in cross-domain scenarios due to the lack of collaborative knowledge from items in cross-domain.
% \textbf{3)} The modality-aware model (MoRec) outperformed CF-RecSys by leveraging textual information from cross-domain items through the modality encoder.

% \noindent\textbf{Cross-domain Scenarios}

% \vspace{-2ex}
\subsection{Ablation Studies}
In this section, we show ablation studies for our model. We mainly analyze 
the effect of each component in~\proposed~regarding Stage-1 (Section~\ref{exp: ablation for losses in stage1}) and Stage-2 (Section ~\ref{exp: ablation for stage2}).
% and the integration of collaborative knowledge on LLM in Stage-2 (Section~\ref{Sec phase2}) in Section ~\ref{exp: ablation for stage2}.
\vspace{-2ex}
\begin{table}[h]
  \caption{Ablation studies on Stage-1 of~\proposed~(Hit@1).}
  \label{tab:ablation 1}
    \vspace{-2ex}
\resizebox{0.9\linewidth}{!}{
\begin{tabular}{c|c|c|c}
\toprule
             Ablation    & Movies and TV & Beauty & Toys \\ \midrule\midrule
\proposed                                               & \textbf{0.6237}    & \textbf{0.5809}    & \textbf{0.3336} \\ \midrule
w/o $\mathcal{L}_\text{matching}$                               & 0.5838    & 0.5548    & 0.3225  \\ \midrule
w/o $\mathcal{L}_\text{item-recon} \& \mathcal{L}_\text{text-recon}$    & 0.5482    & 0.5327    & 0.3204 \\ \midrule
w/o $\mathcal{L}_\text{rec}$                           &  0.6130   & 0.5523    & 0.1541      \\  \midrule
Freeze SBERT  &  0.6173 & 0.5565 & 0.1720 \\ 
\bottomrule
\end{tabular}}
    \vspace{-3ex}
\end{table}

\subsubsection{Effect of Components in Stage-1}
\label{exp: ablation for losses in stage1}
This section presents the experimental results showing the benefit of each component during the Stage-1. 
% Our ablation studies utilized three datasets.
Across all datasets, the exclusion of any loss resulted in decreased performance. We make the following observations:
\textbf{1)} Removing $\mathcal{L}_\text{matching}$ from in Equation ~\ref{Eq matching loss} results in a significant performance decline across all datasets. 
% Specifically, omitting the $\mathcal{L}_{match}$ (Equation ~\ref{Eq matching loss}) led to a performance decline of up to approximately 4 percentage points.
This implies that the alignment between the item and the text information is effective and that the LLM can comprehend item textual information in joint collaborative-text embeddings to enhance recommendation capabilities.
\textbf{2)} Removing $\mathcal{L}_\text{item-recon}$ and $\mathcal{L}_\text{text-recon}$ leads to performance drop, owing to the risk of over-smoothed representations (i.e., $\mathbf{e}\approx \mathbf{q}$), as discussed in Section ~\ref{sec: reconstruction}. 
% This implies that the reconstruction losses effectively avoids over-smoothed representations of the joint collaborative-text representations, allowing us to retain the unique information within the joint representaitons.
% Another set of experiments involved the exclusion of reconstruction losses. In stage one, we applied two types of reconstruction losses: one for item embedding and another for text embedding. We have previously discussed the risk of over-smoothing between embeddings, which can empirically reduce performance.
\textbf{3)} We observe that removing $\mathcal{L}_\text{rec}$ leads to performance drop. Since $\mathcal{L}_{rec}$ is introduced to explicitly incorporate the collaborative knowledge while informing the model about the recommendation task, the performance drop indicates the reduction of collaborative knowledge between items and users, which is crucial for recommendation tasks. 
\textbf{4)} Lastly, we kept SBERT frozen while training~\proposed. We observe that freezing SBERT leads to poor performance across all datasets. This implies that fine-tuning SBERT facilitates the text embeddings to adapt to the recommendation task.
% \begin{table}[t]
%   \caption{Ablation studies on Stage-1 of~\proposed~(Hit@1).}
%   \label{tab:ablation 1}
%     \vspace{-2ex}
% \resizebox{0.72\linewidth}{!}{
% \begin{tabular}{c|c|c|c}
% \toprule
%              Ablation    & Movies and TV & Beauty & Toys \\ \midrule\midrule
% \proposed                                               & \textbf{0.6237}    & \textbf{0.5809}    & \textbf{0.3336} \\ \midrule
% w/o $\mathcal{L}_\text{matching}$                               & 0.5838    & 0.5548    & 0.3225  \\ \midrule
% w/o $\mathcal{L}_\text{item-recon} \& \mathcal{L}_\text{text-recon}$    & 0.5482    & 0.5327    & 0.3204 \\ \midrule
% w/o $\mathcal{L}_\text{rec}$                           &  0.6130   & 0.5523    & 0.1541      \\  \midrule
% Freeze SBERT  &  0.6173 & 0.5565 & 0.1720 \\ 
% \bottomrule
% \end{tabular}}
%     \vspace{-3ex}
% \end{table}

\vspace{-1.5ex}
\begin{table}[h]
  \caption{Ablation study on Stage-2 of~\proposed~(Hit@1).}
  \label{tab:ablation 2}
      \vspace{-2ex}
\resizebox{1.0\linewidth}{!}{
\begin{tabular}{c|c|cccc}
\toprule
Row &  Ablation   & Movies and TV & Video Games & Beauty & Toys   \\ \midrule\midrule
(1) &\proposed      & \textbf{0.6237} & \textbf{0.5282} & \textbf{0.5809} & \textbf{0.3336} \\ \midrule
(2) &\proposed~w/o user representation & 0.5925 & 0.5121 & 0.5547 & 0.3217  \\ \midrule
(3)& \proposed~w/o joint embedding  &0.1224 & 0.4773 & 0.5213 & 0.2831 \\ \midrule
(4)& \proposed~with random joint embedding &0.1200 & 0.4729 & 0.5427 & 0.0776 \\ \bottomrule
\end{tabular}}
    \vspace{-2ex}
\end{table}

\subsubsection{Effect of the Alignment method in Stage-2}
\label{exp: ablation for stage2}
% item embedding, user rep 모두 recommendation에 도움이 되는 정보를 담고 있다.
% item embedding이 collaborative knowledge에 대한 정보가 더 많이 담고 있기 때문에 더 큰 영향을 준다.
    % + 아이템에 대한 text information이 담겨있음
% Un-pretrained RecSys를 합쳤을 때, collaborative knowledge가 넘어가지 않는 상황
    % => TallRec과 비슷하거나 낮아지는 상황
    % => 
% User and item embeddings, which encapsulate collaborative knowledge, are transferred and aligned to the LLMs.
% This section presents the experimental results showing the benefit of each component during the Stage-2. 
Recall that a user representation and item embeddings are injected to the LLM prompt as shown in Figure~\ref{fig: prompt}. In this section, we verify the benefit of injecting them into the prompt (rows (2-4) in Table ~\ref{tab:ablation 2}). We have the following observations:
% The ablation study is divided into four scenarios: one without user embeddings, one without joint embeddings, one with randomly initialized item embeddings, and one with joint embeddings from SBERT.
Across all datasets, 
\textbf{1)} the absences of either the user representation (row (2)) or the joint embedding (row (3)) from the prompt led to a reduction in performance. Notably, the exclusion of the joint embedding results in a more substantial decrease, underscoring its significant role in transferring collaborative knowledge. Moreover, as joint embeddings also capture the textual information about items, their exclusion is particularly detrimental.
% While user embeddings appear to be less vital than joint embeddings, they still contribute positively to performance.
\textbf{2)} When we replace the joint embedding with a randomly initialized embedding (row (4)), which means~\proposed~is trained with item embeddings without collaborative knowledge,
we observe performance degradation across all datasets. This indicates the importance of leveraging the collaborative knowledge for recommendation.
% , with Movies and TV, and Toys categories experiencing the most pronounced declines. It is worth mentioning that the performance trends observed were in line with those of TALLRec in Table~\ref{tab: main table}, suggesting a pattern consistent with the underlying collaborative information alignment to LLM, and emphasizing the importance of imparting collaborative information to LLM.
% \textbf{3)} Lastly, we observe performance drop when using $\mathbf{q}$ instead of $\mathbf{e}$ for training~\proposed~(row (5)) from Stage-1. This can be attributed to the change in the source of the joint embeddings, i.e., from CF-RecSys to SBERT, which results in a decrease in collaborative knowledge within the joint embeddings. It is important to note that row (5) is different from~\proposed~(SBERT) that we compared throughout the experiments in that \proposed~(SBERT) is trained with $\mathbf{e}$ while the inference is conducted based on $\mathbf{q}$, whereas row (5) is trained with $\mathbf{q}$ and inference is conducted on $\mathbf{q}$.
% As a result, there is an overall decline in performance, which contrasts with the observation in cold item scenarios discussed in Section \ref{exp: cold/warm item}, where the utilization of textual information played a pivotal role over collaborative information.

% 추가적인 insight 혹은 analysis?

\subsection{Model Analysis}
\subsubsection{Train/Inference Speed}
\label{exp: model_speed}
Recall that \proposed~{requires the fine-tuning of neither the CF-RecSys nor the LLM}. Specifically,~\proposed~efficient in that the alignment network is the only trainable neural network, while TALLRec~\cite{bao2023tallrec} requires the fine-tuning of the LLM with LoRA.
% and the alignment network is the only neural network that is trained in~\proposed. 
% ~\proposed~only aligns the intermediate network, requiring less cost and time compared to the previous LLM-based model, TALLRec. This simpler model architecture offers the advantage of faster training and inference times. 
In this section, we compare the training and the inference time of~\proposed~and TALLRec.
% also learning under the auto-regressive manner of ~\proposed~ as mentioned in Section \ref{Sec phase1}, using the Amazon Beauty dataset. 
As for the training time, we measured the total time spent until the end of training, and as for the inference time, we measured the time spent per mini-batch. Table \ref{tab: model speed} shows that~\proposed~exhibits significantly faster training and inference time compared with TALLRec. Notably, a more substantial improvement is observed in training time, since~\proposed~does not require the LLM to be fine-tuned unlike TALLRec, which demonstrates the applicability of LLM in large-scale recommendation datasets. Moreover, the faster inference time demonstrates the practicality of~\proposed~in real-world scenarios, especially in the context of real-time recommendation services where inference time is critically important.
% Notably, the improvement in inference speed is more significant. Since generative base models require time for inference, reducing this time suggests a higher feasibility for real-world scenarios. 

% \smallskip
\subsubsection{Training with all items in each sequence. } \label{sec:autoregressive}
Recall that for efficiency in training, we used only the last item of each user sequence when optimizing the final loss in Stage-1 (Equation~\ref{Eq overall loss}) and Stage-2 (Equation~\ref{LLM Loss}) of~\proposed. In this section, we report the recommendation performance in terms of Hit@1 and train/inference speed when using all items in each user sequence for optimization (see \proposed$_\textsf{all}$ in Table \ref{tab: model speed}).
% Additionally, we compared the time and performance when we trained the ~\proposed~ under the auto-regressive manner of ~\proposed~ as mentioned in Section \ref{Sec phase1}. 
We observe that as expected the recommendation performance is further improved when using all items in each user sequence. However, considering that the training time also increased approximately 3 times, the improvement seems marginal.
% Although the learning time of about 3 times that of our ~\proposed, there is a small gap compared to the model learned with auto-regressive methods. 
It is important to note that since vanilla~\proposed~is trained based on only the last item in each user sequence, there is a large amount of unseen/new items that appear in the test set\footnote{About 13\% of items are unseen during training in the Beauty dataset.}. However, valilla~\proposed~still showed comparable performance with~\proposed$_\textsf{all}$, implying the generalization ability of~\proposed.

% our model was not trained auto-regressively, there were unlearned items, but the gap was filled through generalization ability.

\begin{table}[t]
    \caption{Train/Inference time comparison (Beauty dataset).}
    \label{tab: model speed}
        \vspace{-2ex}
    \resizebox{0.73\linewidth}{!}{
    \begin{tabular}{c|cc|c}
    \toprule
                & Train time (min)    & Inference time (sec/batch) & Hit@1 \\ \midrule\midrule
    TALLRec     & 588.58              & 3.36      &  0.5542    \\ \midrule
    \proposed   & \textbf{232.5}    & \textbf{1.98}   & 0.5809 \\ \midrule
    \proposed$_\textsf{all}$  &  {643.33} & {\textbf{1.98}} & {\textbf{0.6002}} \\ \bottomrule
    \end{tabular}}
        \vspace{-1ex}
\end{table}

\begin{table}[t]
\caption{Results showing~\proposed~is model-agnostic.}
  \label{tab: model agnostic}
    \vspace{-2ex}
    \resizebox{0.5\linewidth}{!}{
\begin{tabular}{c|cc}
\toprule
                 Model  & Beauty & Toys   \\ \midrule\midrule
SASRec               & 0.5298 & 0.2359 \\ \midrule
\proposed~ (SASRec)    & \textbf{0.5809} & \textbf{0.3336} \\ \midrule
NextItNet            & 0.4231 & 0.1415 \\ \midrule
\proposed~ (NextItNet) & 0.5642 & 0.3203 \\ \midrule
GRU4Rec             & 0.4131 & 0.1673  \\ \midrule
\proposed~ (GRU4Rec)  &  0.5542 & 0.3089 \\ \midrule
NCF                 &  0.2957 & 0.1849 \\ \midrule
\proposed~ (NCF)      &  0.5431 & 0.3263 \\ \bottomrule
\end{tabular}}
\vspace{-3ex}
\end{table}

\subsubsection{\proposed~is Model-Agnostic}
\label{exp: model_agnostic}
Although~\proposed~adopts SASRec as the backbone CF-RecSys, it can be replaced with any existing collaborative filtering recommender systems, thanks to the model-agnostic property. 
% sequential recommender 뿐만 아니라 NCF 에서도 잘 된다는 점
% better collaborative knowledge는 ours의 성능을 높여준다.
% => SOTA의 변화에 따라 ours를 적용할 수 있음.
% Our framework ~\proposed~ aligns and trains only the intermediate network, allowing for integration with any existing CF-RecSys. In other words, we can easily select a SOTA model in the domain of recommendations. 
Hence, we adopt three other collaborative filtering recommender systems including two sequential recommenders (i.e., NextItNet and GRU4Rec), and one non-sequential recommender (i.e., NCF) to~\proposed.
We make the following observations from Table \ref{tab: model agnostic}.
\textbf{1)} Adopting the SASRec backbone performs the best, which is expected since SASRec outperforms other CF-RecSys in their vanilla versions.
This implies that transferring high-quality collaborative knowledge can enhance the performance of~\proposed. 
\textbf{2)} Adopting~\proposed~to any backbone improves the performance of the vanilla model. This implies that if the SOTA model changes in the future, our framework has the potential to further improve performance by replacing the existing CF-RecSys in the framework.
\textbf{3)} We observe that while the performance difference between SASRec and NCF is nearly double when they operate as standalone CF-RecSys, the integration with ~\proposed, which leverages the modality of item text information and the intensive capabilities of LLM, reduces this performance gap.
% Table \ref{tab: model agnostic} indicates that our framework is CF-RecSys model-agnostic not only sequential recommender but also MF-based models. And remarkable point is that when ~\proposed~ integrated with SASRec, it achieves the highest performance while the SASRec is the most superior model (SOTA model) among collaborative-based models. This implies that transferring high-quality collaborative knowledge can enhance the performance of our proposed framework. Furthermore, should the SOTA model change in the future, our framework has the potential to exhibit even better performance than before by replacing the CF-RecSys incorporated in the framework.
% We also observe that while the performance difference between SASRec and NCF is nearly double when they operate as standalone CF-RecSys, the integration with our proposed framework, which leverages the modality of item text information and the intensive capabilities of LLM, can reduce this performance gap.

\begin{figure}[h]
    \centering
    \vspace{-1ex}
    \includegraphics[width=1\linewidth]{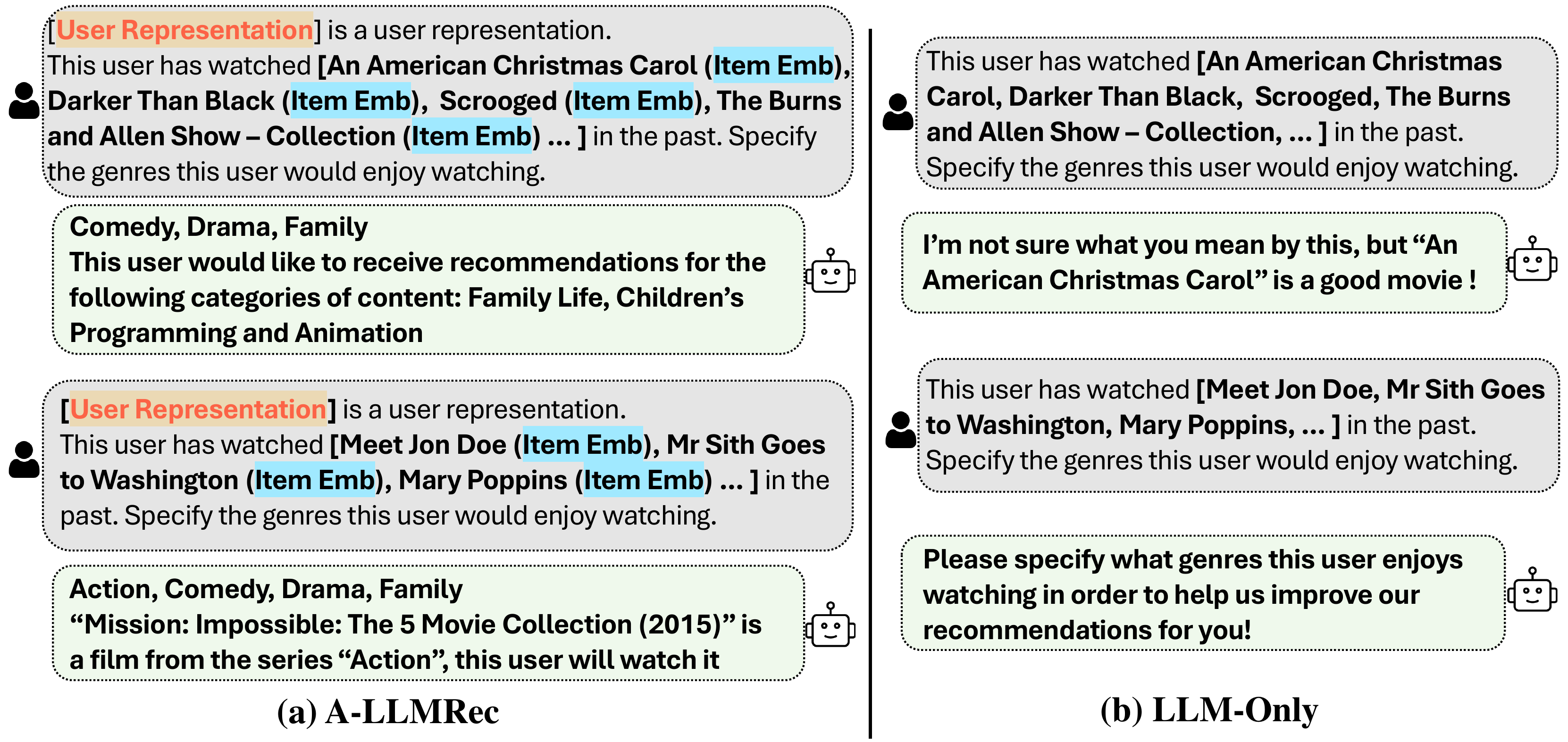}
    \vspace{-1ex}
    \caption{\proposed~v.s. LLM-Only on the favorite genre prediction task (Movies and TV dataset used).
    }
    \label{fig: genre generation}
    \vspace{-3ex}
\end{figure}

\subsubsection{Beyond Recommendation: Language Generation Task (Favorite genre prediction)}
\label{exp: Genre generation}
% Since~\proposed~
To validate whether~\proposed~can generate natural language outputs based on the understanding of users and items
through the aligned collaborative knowledge from CF-RecSys, we conduct a favorite genre prediction task (Figure~\ref{fig: genre generation}). That is, given the same prompt format, we ask the LLM-based models (i.e., \proposed~and LLM-Only) using the same backbone LLM, which is OPT-6.7B, to predict the movie genres that a given user would enjoy watching. 
The only difference in the prompt is that while LLM-only is only given titles of movies watched by the user in the past,~\proposed~is given the user representation and item embeddings along with the movie titles. In Figure~\ref{fig: genre generation}, we observe that~\proposed~indeed generates proper answers, while LLM-Only fails to do so.
We attribute this to the fact that the item embeddings of the CF-RecSys are well aligned with the token space of the LLM, which enables the LLM to understand and utilize collaborative knowledge.
Note that although we also experimented with TALLRec, we were not able to obtain valid outputs. 
% We conjecture that since the LLM in TALLRec is fine-tuned via instruction tuning process that makes the model to respond with a binary answer of ``Yes'' or ``No'', generating valid natural language outputs has become a non-trivial task.
We conjecture that since the LLM in TALLRec is fine-tuned via an instruction-tuning process that makes the model provide responses as part of the recommendation task, generating valid natural language outputs has become a non-trivial task.
Please refer to {{Appendix \ref{append: generation}}} for the results of TALLRec.

\section{Conclusion}
% outperform across various scenarios
% Model agnostic, Time efficiency
In this paper, we propose a novel LLM-based recommender system, named ~\proposed. The main idea is to enable LLMs to utilize the collaborative knowledge from pre-trained CF-RecSys. By doing so, \proposed~outperforms existing CF-RecSys, modality-aware recommender systems, and LLM-based recommenders under various scenarios including cold/warm items, cold user, few-shot, and cross-domain scenarios.
Moreover, we also demonstrate that the two advantages originated from fine-tuning neither pre-trained CF-RecSys nor LLMs, i,e, Model-agnostic and efficiency. 
% Specifically, \proposed~can be easily integrated with any existing CF-RecSys, and since the alignment network is the only trainable network, \proposed~is efficient in both at train and inference.
{Lastly, we show the potential of A-LLMRec in generating natural language tasks based on the understanding of collaborative knowledge from CF-RecSys.} For future work, we plan to further enhance the ability of the LLM in~\proposed~based on advanced prompt engineering such as chain-of-thought prompting~\cite{wei2022chain}.

% \section{Ethics Statement}
\smallskip \noindent \textbf{Ethics Statement}
To the best of our knowledge, this paper aligns with the KDD Code of Ethics without any ethical concerns. The datasets and codes employed in our research are publicly available.

% \smallskip \noindent \textbf{Acknowledgements} 
\begin{acks}
This work was supported by NAVER Corporation, the National Research Foundation of Korea(NRF) grant funded by the Korea government(MSIT) (RS-2024-00335098), and National Research Foundation of Korea(NRF) funded by Ministry of Science and ICT (NRF-2022M3J6A1063021).
\end{acks}

% \section{SIGCHI Extended Abstracts}

% The ``\verb|sigchi-a|'' template style (available only in \LaTeX\ and
% not in Word) produces a landscape-orientation formatted article, with
% a wide left margin. Three environments are available for use with the
% ``\verb|sigchi-a|'' template style, and produce formatted output in
% the margin:
% \begin{description}
% \item[\texttt{sidebar}:]  Place formatted text in the margin.
% \item[\texttt{marginfigure}:] Place a figure in the margin.
% \item[\texttt{margintable}:] Place a table in the margin.
% \end{description}


% acknowledgments
% \begin{acks}
% \end{acks}
% \clearpage

\clearpage
\bibliographystyle{ACM-Reference-Format}
\bibliography{sample-base}

\clearpage
\begin{figure*}[t]
    \centering
    \vspace{-1ex}
    \includegraphics[width=0.73\linewidth]{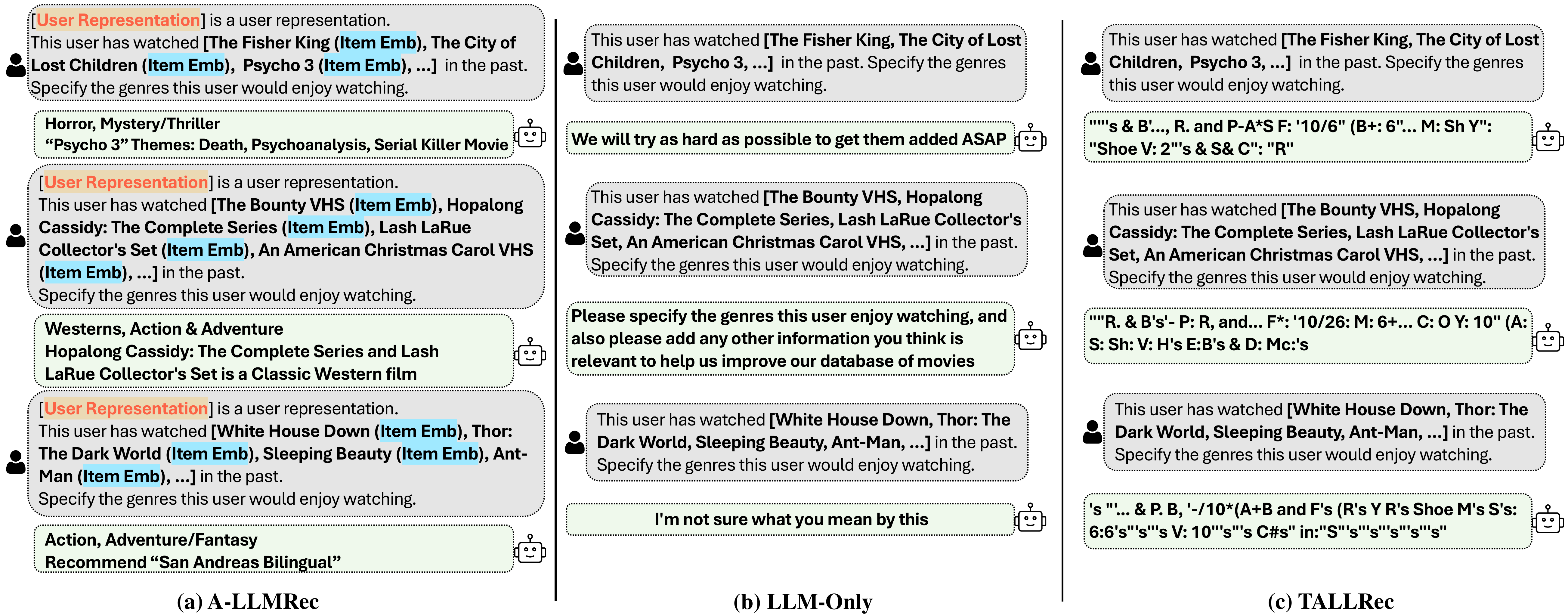}
    \vspace{-2.5ex}
    \caption{\proposed, LLM-Only, and TALLRec on the favorite genre prediction task (Movies and TV dataset used).
    }
    \label{fig: genre appendix}
    \vspace{-3ex}
\end{figure*}

\appendix

% \section{Ethics Statement}
% To the best of our knowledge, this paper aligns with the KDD Code of Ethics without any ethical concerns. The datasets and codes employed in our research are publicly available.

\vspace{-2.8ex}
\section{Baselines}
\label{append: baselines}
\begin{figure}[t]
    \centering
    \includegraphics[width=0.75\linewidth]{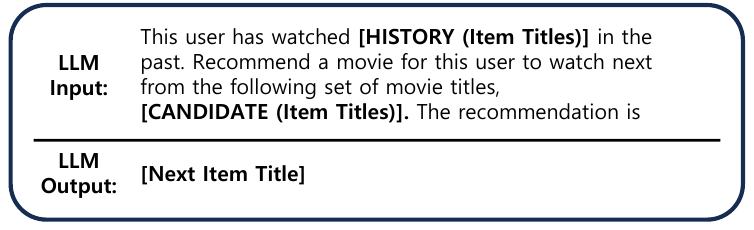}
    \vspace{-2.5ex}
    \caption{An example prompt designed for the Amazon Movies dataset used by LLM-based models, i.e., TALLRec and LLM-Only models.}
    \label{prompt_llm_based}
    \vspace{-3ex}
\end{figure}

\begin{enumerate}[leftmargin=0.5cm]
    \item Collaborative filtering recommender systems
    \begin{itemize}[leftmargin=0.1cm]
        \item \textbf{NCF}~\cite{he2017neural} combines neural networks (MLP) to capture the collaborative information. Note that NCF is a two-tower model comprised of separate components for the user and item embedding matrix.

        \item \textbf{NextItNet}~\cite{yuan2019simple}
        % proposes a temporal convolutional network that utilizes 1D-dilated convolutional layers to capture the long-term dependencies inherent in interaction sequence, and by utilizing residual connections, NextItNet enhances optimization efficiency.
        proposes a temporal convolutional network that utilizes 1D-dilated convolutional layers and residual connections to capture the long-term dependencies inherent in interaction sequence.
        \item \textbf{GRU4Rec}~\cite{hidasi2015session}
        adopts RNNs to model user behavior sequences for session-based recommendations.
        
        \item \textbf{SASRec}~\cite{kang2018self}
        % is a representative ID-based recommender that adopts a self-attention encoding method to model the user preferences from the interaction sequence.
        is our main baseline, a state-of-the-art collaborative filtering recommender system (CF-RecSys) that adopts a self-attention encoding method to model user preferences from user behavior sequences.
    \end{itemize}
    
    \item Modality-aware recommender systems
    \begin{itemize}[leftmargin=0.2cm]
        \item \textbf{MoRec}~\cite{10.1145/3539618.3591932}
        employs a pre-trained SBERT to utilize the text information of items to generate the initial embeddings for items that will be used in collaborative filtering models. We utilize SASRec as the backbone model of MoRec.
        % employs pre-trained BERT to utilize the item text information and enhance the sequential recommendation.
        
        \item \textbf{CTRL}~\cite{li2023ctrl}
        % considers user tabular data and its textual representation and uses them to pre-train collaborative recommendation models.
        employs a two-stage learning process: the first stage involves contrastive learning on textual information of items to initialize the backbone model, and the second stage, fine-tunes the model on recommendation tasks. We use SASRec as the backbone model of CTRL.

        \item \textbf{RECFORMER}~\cite{10.1145/3580305.3599519}
        {models user preferences and item features using the Transformer architecture, transforming sequential recommendation into a task of predicting the next item as if predicting the next sentence, by converting item attributes into a sentence format.}
    \end{itemize}
    
    \item LLM-based recommender systems
    \begin{itemize}[leftmargin=0.2cm]
        \item \textbf{LLM-Only}
        utilizes an open-source LLM model OPT~\cite{zhang2022opt} with prompts related to recommendation tasks as shown in Figure~\ref{prompt_llm_based}. In our experiments, we adopt the 6.7B size version of OPT for all LLM-based recommendations.
        
        \item \textbf{TALLRec}~\cite{bao2023tallrec} is our main baseline, which learns the recommendation task based on prompts consisting solely of text and fine-tunes the LLMs using the LoRA. 
        Their approach involves providing user interaction history and one target item and determining whether a user will prefer this target item. This simpler task necessitates only a brief prompt for the LLMs.
        % The authors can set a maximum input length of 512 and a batch size of 128.
        In contrast, our recommendation task requires a more extensive prompt.
        % , more than twice as long, as it must include 20 candidate items. Consequently, we have disabled the truncation mode for the LLM and removed the maximum length constraint,
        Even though this adjustment results in a smaller batch size, the same as ~\proposed, for training TALLRec. We use the prompt shown in Figure~\ref{prompt_llm_based}.

        \item \textbf{MLP-LLM}
        is an additionally designed LLM-based recommendation model for analysis. Compared with~\proposed, this model directly connects the user and item embeddings from frozen CF-RecSys and LLM using only MLP layers, instead of the auto-encoders in~\proposed~that involve various techniques to align the collaborative knowledge of CF-RecSys with the LLM. Note that we use the prompt shown in Figure~\ref{fig: prompt}.
    \end{itemize}
    
\end{enumerate}

\begin{table}[]
    \small
    \centering
    \caption{Source code links of the baseline methods.}
    \vspace{-2ex}
    \resizebox{0.8\linewidth}{!}{
    \begin{tabular}{c|c}
    Methods & Source code \\ \hline \hline
    SASRec & \url{https://github.com/pmixer/SASRec.pytorch} \\
    NextItNet & \url{https://github.com/syiswell/NextItNet-Pytorch} \\
    GRU4Rec & \url{https://github.com/hungpthanh/GRU4REC-pytorch} \\
    RECFORMER & \url{https://github.com/AaronHeee/RecFormer}\\
    TALLRec & \url{https://github.com/SAI990323/TALLRec}\\
    \hline
    \proposed & \url{https://github.com/ghdtjr/A-LLMRec} \\
    \hline
    \end{tabular}}
    \label{tab: codes}
    \vspace{-4ex}
\end{table}

% \section{Hyperparamter Analysis}
% \label{append: hyper}
% {In Table \ref{tab: sensitivity} we conduct an analysis regarding the hyperparameter sensitivity on the Beauty dataset to verify the robustness of~\proposed~over hyperparameters, especially on $\alpha$ and $\beta$ in Equation \ref{Eq overall loss}. In Table ~\ref{tab: sensitivity}, although the performance of~\proposed~is the best when $\alpha=0.5$ and $\beta=0.2$,~\proposed~still outperforms TALLRec without carefully setting the hyperparameters. This implies the superiority of~\proposed.
% % As demonstrated in Table ~\ref{tab: sensitivity}, when trained without carefully adjusted hyperparameters like setting 1 or 0.5 for both parameters, \proposed outperforms the baselines including TALLRec, for the Beauty Dataset.
% }

\vspace{-2.3ex}
\section{Language Generation Task}
\label{append: generation}
In Figure \ref{fig: genre appendix}, we present additional favorite genre prediction task results for experiment in shown in Section \ref{exp: Genre generation}. As mentioned in Section \ref{exp: Genre generation}, TALLRec could not generate valid natural language outputs due to the fine-tuning via instruction tuning process, which makes the LLM of TALLRec being able to answer only with some particular prompts used in instruction tuning process. The additional results indicate that~\proposed~can generate the favorite genres for the users based on the understanding of the aligned user representation and item embeddings while LLM-only fails to do so.

\vspace{-1.5ex}
\section{Reproducibility}
\label{append: Reproducibility}
For implementing the baseline, we followed the official codes published by authors as detailed in Table \ref{tab: codes}.
% We conduct experiments within the same environment.
Refer to our source code and instructions to run code for reproducing the results reported in the experiments.

\end{document}